\documentclass[%
 reprint,
superscriptaddress,
 amsmath,amssymb,
 aps,
prb,
]{revtex4-2}

\usepackage{graphicx}
\usepackage{dcolumn}
\usepackage{bm}
\usepackage[english]{babel} 
\usepackage{hyperref}
\usepackage{physics}
\usepackage{amsmath}
\usepackage{subfigure}
\usepackage{bm}
\usepackage{epstopdf}

\begin{document}

\title{Spin-polarized hot electron transport versus spin pumping mediated by local heating}

\author{M. Beens}
\email[Corresponding author: ]{m.beens@tue.nl}
\affiliation{Department of Applied Physics, Eindhoven University of Technology \\ P.O. Box 513, 5600 MB Eindhoven, The Netherlands}
\author{K.A. de Mare}
\affiliation{Department of Applied Physics, Eindhoven University of Technology \\ P.O. Box 513, 5600 MB Eindhoven, The Netherlands}
\author{R.A. Duine }
\affiliation{Department of Applied Physics, Eindhoven University of Technology \\ P.O. Box 513, 5600 MB Eindhoven, The Netherlands}
\affiliation{ Institute for Theoretical Physics, Utrecht University \\
Leuvenlaan 4, 3584 CE Utrecht, The Netherlands}
\author{B. Koopmans}
\affiliation{Department of Applied Physics, Eindhoven University of Technology \\ P.O. Box 513, 5600 MB Eindhoven, The Netherlands}

\date{\today}

\begin{abstract}
A `toy model'\textemdash aimed at capturing the essential physics\textemdash is presented that jointly describes spin-polarized hot electron transport and spin pumping driven by local heating. These two processes both contribute to spin-current generation in laser-excited magnetic heterostructures. The model is used to compare the two contributions directly. The spin-polarized hot electron current is modeled as one generation of hot electrons with a spin-dependent excitation and  relaxation scheme.  Upon decay, the excess energy of the hot electrons is transferred  to a thermalized electron bath. The elevated electron temperature leads to an increased rate of electron-magnon scattering processes and yields a local accumulation of spin. This process is dubbed as spin pumping by local heating. The built-up spin accumulation is effectively driven out of the ferromagnetic system by (interfacial) electron transport. Within our model, the injected spin current is  dominated by the contribution resulting from spin pumping, while the hot electron spin current remains relatively small. We derive that this observation is related to the ratio between the Fermi temperature and Curie temperature, and we show what other fundamental parameters play a role.
\end{abstract}

\maketitle

\section{Introduction}

The generation of spin transport by femtosecond laser-pulse excitation paves the way towards ultrafast spintronic applications. Similar to subpicosecond quenching of the magnetization  \cite{Beaurepaire1996,Zhang2000,Koopmans2005,Kazantseva2007,carpene2008dynamics,Bigot2009,Krauss2009,Koopmans2010,Battiato2010,Atxitia2011,Manchon2012,mueller2013feedback,Mueller2014,Haag2014,Nieves2014,Tveten2015,Krieger2015,Tows2015,Dornes2019,Tauchert2021}, the physical origin of laser-induced spin transport is an unsolved quest and remains heavily debated after more than a decade of experimental and theoretical research  \cite{malinowski2008control,Melnikov2011,Battiato2010,Battiato2012,Choi2014,Kampfrath2013,Choi2015,Alekhin2017,Kimling2017,Seifert2018,Iihama2018,Igarashi2020,Remy2020,Beens2020,VanHees2020,Iihama2021,Lichtenberg2022,Rouzegar2021}
Nevertheless, it is clear that the manipulation of magnetic materials with femtosecond laser pulses is unique by being ultrafast and very efficient. Therefore, understanding the underlying physical mechanisms is interesting from both a fundamental and technological viewpoint.  

There are two dominant theories on the physical origin of  ultrafast spin currents. First, the laser pulse generates a population of highly energetic electrons that through spin-dependent excitation rates and mobilities yield a spin-polarized hot electron current. Including the generated cascades of secondary hot electrons, it is an efficient scheme of spin-current generation, as is described by the model for  superdiffusive spin transport \cite{Battiato2010,Battiato2012}. The second theory is based on the notion that laser heating results in an increased rate of spin-flip scattering processes, including electron-magnon scattering. The latter generates a local spin accumulation \cite{Choi2014,Tveten2015,Shin2018}, a process referred to as bulk spin pumping \cite{tserkovnyak2005nonlocal,Shin2018}, that effectively can be transported towards a neighboring nonmagnetic layer through spin diffusion.  With the two major viewpoints in mind, the essential unanswered question is whether the generated spin current is a direct result of the excitation of hot electrons or is indirectly driven by heating and subsequent spin pumping.

In this work, we present a simplified phenomenological model\textemdash also referred to as `toy model'\textemdash that jointly describes the generated hot electron spin currents and the spin currents driven by spin pumping. Hot electron transport is described by one generation of optically-excited electrons with spin-dependent excitation and decay rates. Within our approach, the hot electrons decay into an instantaneously thermalized electron bath, where the absorbed excess energy results in an increase of the electron temperature. The latter, and the coupling to a thermal magnon bath, is calculated explicitly. It gives an expression for the total built-up spin accumulation and the resulting spin current transported by the thermal electrons. The two contributions to the spin current are calculated equivalently at the interface of a ferromagnetic metal/nonmagnetic metal heterostructure. We show that the spin current driven by spin pumping dominates and we derive that this observation is related to the ratio between the Fermi temperature and Curie temperature. Finally, we discuss the presence of spin-polarized screening currents and investigate their role.  

\begin{figure}[t!]
\includegraphics[scale=0.95]{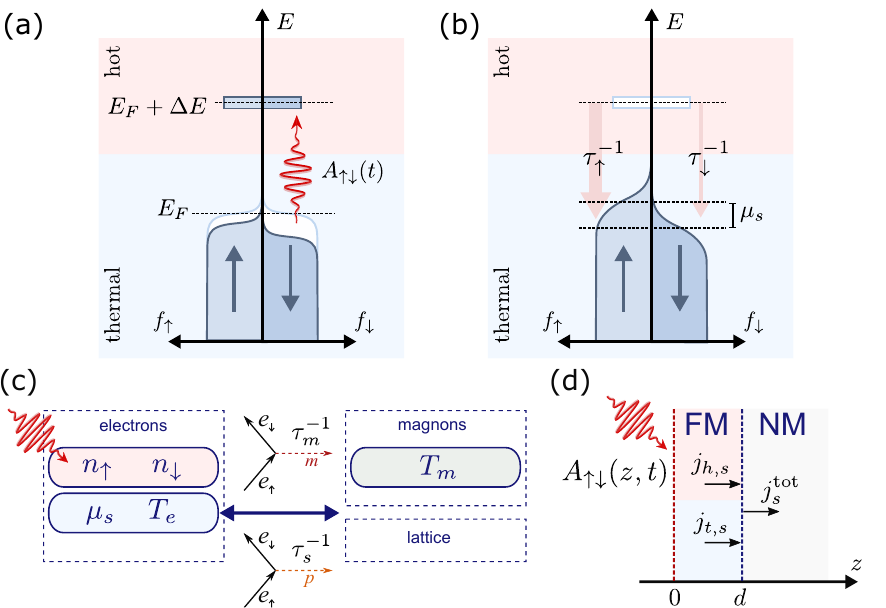}
\caption{\label{fig:5-1} Schematic overview of the toy model. The system of electrons is composed of hot electrons and a bath of electrons that remains thermalized. (a) Excitation by a laser pulse yields a population of hot electrons and a reduction of the number of  electrons in the thermal system. (b) The spin-dependent excitation and decay rates yield a net spin accumulation within the thermal system. (c) The decay  processes are associated with an energy absorption of the thermal system, leading to an increased temperature $T_e$. Subsequent interactions among electrons, magnons, and the lattice, compose the full spin angular momentum flow. (d) Schematic overview of the modeled heterostructure, excited at position $z=0$ by absorption profile $A_{\uparrow\downarrow}(t)$. The total interfacial spin current $j_s^\mathrm{tot}$ is the sum of the hot electron spin current $j_{h,s}$ and the spin current generated within the thermal electron system $j_{t,s}$. 
} 
\end{figure}

\section{Toy model for laser-induced hot electron dynamics} 

We start with defining two categories of electrons for each spin polarization  separately. First, the electrons far above the Fermi level are defined as `hot' electrons. Secondly, the electrons close to and far below the Fermi level are assumed to remain thermalized and are dubbed as `thermal' electrons. We treat the thermal electron system as a single population of mobile electrons, representing hybridized $3d$ and $4s$ electrons in the transition metal ferromagnets, composed of the subsystems for majority spins (here defined as $\downarrow$) and minority spins ($\uparrow$). A schematic overview is given in Figs.\ \ref{fig:5-1}(a)-(b). Upon excitation, a (spin-polarized) population of thermal electrons is transferred to the higher energetic `hot' state at energy $\Delta E$ above the Fermi level. In combination with the spin-dependent decay rates $\tau_{\uparrow\downarrow}^{-1}$, this leads to a shift of the spin-dependent chemical potential $\mu_s=\mu_\uparrow-\mu_\downarrow$. In other words, a spin accumulation is created. During the decay of the hot electrons, the excess energy $\Delta E $ is absorbed by the thermal system and leads effectively to an elevated electronic temperature $T_e$. The latter gives rise to the creation  of thermal magnons and an additional change of the spin accumulation $\mu_s$, as will be discussed in Sec.\ \ref{sec:5-3}. 

\begin{figure}[t!]
\includegraphics[scale=1.00]{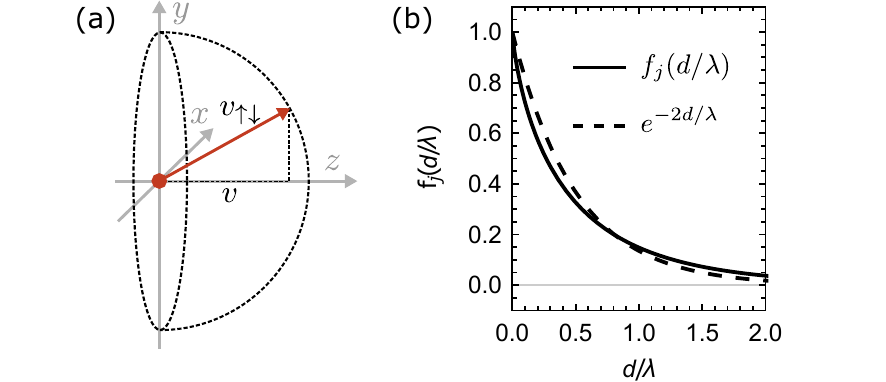}
\caption{\label{fig:5-2} (a) An excited hot electron with spin up ($\uparrow$) or spin down ($\downarrow$) moves in a random direction with speed $v_{\uparrow\downarrow} $ and longitudinal velocity  component $v$.  (b) The function $f_j(d/\lambda)$, which describes the spatial decay of the hot electron spin current, plotted as a function of the normalized thickness $d/\lambda$  (black solid line). The function is compared to the exponent of $-2d/\lambda$, represented by the dashed black line. 
} 
\end{figure}

We first focus on the hot electron transport generated after excitation. We consider a (magnetic) metallic system described by spin-dependent electron distribution functions that remain homogeneous in the transverse plane but may vary along the longitudinal (out-of-plane) $z$ direction. Furthermore, as schematically depicted in Fig.\ \ref{fig:5-2}(a), we assume that when hot electrons are excited they move in a random direction with a (spin-dependent) fixed speed $v_{\uparrow\downarrow}$ until they decay after time $\tau_{\uparrow\downarrow}$. The distribution function describing this hot electron system satisfies the Boltzmann equation 
\begin{equation}
\label{eq:5-1}
\dfrac{\partial n_{\uparrow\downarrow}(z,v,t)}{\partial t} 
+ v \dfrac{\partial n_{\uparrow\downarrow}(z,v,t) }{\partial z}
=
A_{\uparrow\downarrow} (z,t) -\dfrac{n_{\uparrow\downarrow}(z,v,t)}{\tau_{\uparrow\downarrow}} ,
\end{equation}
where $n_{\uparrow\downarrow}(z,v,t)$ corresponds to the distribution function for hot electrons with up ($\uparrow$) and down ($\downarrow$) spin at position $z$ with velocity component $v$ along the $z$ axis. The function $A_{\uparrow\downarrow}(z,t)$ describes the spatiotemporal profile of the laser-pulse absorption and is spin-dependent due to the different absorption coefficients for up and down spins. For simplicity, we assume that this source term is a Dirac delta function located at $z=0$, having $A_{\uparrow\downarrow}(z,t)=A_{0,\uparrow\downarrow} (t) \delta (z)$ (see Fig.\ \ref{fig:5-1}(d)), where $A_{0,\uparrow\downarrow} (t)$ is determined by the temporal profile of the laser pulse. Only focusing on this simplified example is relevant, since the response to a general spatial-dependent function can be calculated straightforwardly by performing a convolution \cite{Battiato2012}. Furthermore, we define the polarization coefficient $P_A=(A_{0,\uparrow}(0)-A_{0,\downarrow}(0))/(A_{0,\uparrow}(0)+A_{0,\downarrow}(0))$ such that $A_{0,\uparrow\downarrow}(t) = A_0(t)(1\pm P_A)$, where $A_0(t)$ corresponds to the spin-averaged excitation profile.    

Using Fourier transformation, we switch from the time domain to the frequency domain, which simplifies the following calculations because convolutions now correspond to a multiplication. We are interested in the dynamics in the region $z\geq 0$, where the solution to Eq.\ (\ref{eq:5-1}) is given by 
\begin{equation}
\label{eq:5-2}
n_{\uparrow\downarrow}(z,v,\omega) =
\dfrac{A_0(\omega)(1\pm P_A) }{v}  \exp(-\dfrac{z}{v\tau_{\uparrow\downarrow}}(1+i\omega \tau_{\uparrow\downarrow}) )  \theta(v).
\end{equation}
The Heaviside theta function $\theta(v)$ makes sure that the solution does not diverge in the limit $z\rightarrow \infty $, meaning that only right-moving electrons are present. We assume that all hot electrons move in a random (positive) direction with a fixed speed $v_{\uparrow\downarrow}$. The number density of the electrons can then be written as 
\begin{equation}
\label{eq:5-3} 
n_{\uparrow\downarrow}(z,\omega) =
\dfrac{A_0(\omega)(1\pm P_A)}{v_{\uparrow\downarrow} }  f_n\Big[\dfrac{z(1+i\omega \tau_{\uparrow\downarrow})}{\lambda_{\uparrow\downarrow} } \Big]  ,
\end{equation}
where the function $f_n(x)$ results from a surface integral over a positive hemisphere with radius $v_{\uparrow\downarrow}$ and we used $\lambda_{\uparrow\downarrow}=v_{\uparrow\downarrow}\tau_{\uparrow\downarrow}$. The proper normalization factors are defined within $A_0(\omega) $. Similarly, the current densities can be expressed as 

\begin{equation}
\label{eq:5-4}
j_{\uparrow\downarrow}(z,\omega) =
A_0(\omega)(1\pm P_A)\, f_j\Big[\dfrac{ z }{\lambda_{\uparrow\downarrow} }(1+i\omega \tau_{\uparrow\downarrow}) \Big]  . 
\end{equation} 

Since the solutions  follow from the Boltzmann equation, the functions $f_n(x)$ and $f_j(x)$ satisfy $f'_j(x) = -f_n(x)$. The function $f_j(x)$ is plotted in Fig.\ \ref{fig:5-2}(b),  showing its similarity with exponential decay. Keeping the latter in mind, the inverse Fourier transform of Eq.\ (\ref{eq:5-4}) approximately corresponds to an exponential decay with length scale $\lambda_{\uparrow\downarrow}/2$ and a phase shift  $2 z /v_{\uparrow\downarrow}$ compared to the temporal profile of the laser pulse.   

Although we focus on a magnetic heterostructure in the following paragraphs, we assume for hot electron transport that the system is homogeneous, since we aim for a simple toy model. The hot electron current at the interface of the heterostructure is simply assumed to be equal to Eq.\ (\ref{eq:5-4}) being evaluated at $z=d$, where $d$ is the thickness of the (imaginary) ferromagnetic layer. The interfacial hot electron spin current $j_{h,s}=j_\uparrow-j_\downarrow $ is included in Fig.\ \ref{fig:5-5}. The figure presents a schematic overview of all contributions to the interfacial spin current. The remaining terms, which mainly represent spin-current contributions in the thermal electron system (indicated by the blue shaded region), will be step-by-step introduced in the following sections.

As Fig.\ \ref{fig:5-5} indicates, for determining the spin transport in the thermal electron system it is required to calculate the functions that characterize the (spin-dependent)  decay of hot electrons. In order to do so, we need the spatial average of $n_{\uparrow\downarrow}(z,\omega)$ over the domain $(0,d]$, notated simply as $n_{\uparrow\downarrow}(\omega)$, which is given by 
\begin{eqnarray}
\label{eq:5-5}
n_{\uparrow\downarrow}(\omega) 
&=& \dfrac{A_0(\omega)(1\pm P_A)}{v_{\uparrow\downarrow}} 
\dfrac{1}{1+i\omega \tau_{\uparrow\downarrow} } 
\dfrac{\lambda_{\uparrow\downarrow}}{d} 
\\
\nonumber
&&\times   \bigg(1-f_j\Big[\dfrac{z}{\lambda_{\uparrow\downarrow} }(1+i\omega \tau_{\uparrow\downarrow}) \Big] \bigg),
\end{eqnarray} 
using the relation between $f_n(x)$ and $f_j(x)$. For convenience, we define one more function that will become relevant in the second part of this article 
\begin{eqnarray}
\label{eq:5-6}
\nonumber 
F_{\pm}(\omega) &=&\pm 
(1+P_A)
\dfrac{1-f_j\Big[\dfrac{d}{\lambda_{\uparrow} }(1+i\omega \tau_{\uparrow}) \Big] }{1+i\omega \tau_\uparrow} 
\\ 
 &&
 + (1-P_A)\dfrac{1-f_j\Big[\dfrac{d}{\lambda_{\downarrow} }(1+i\omega \tau_{\downarrow}) \Big] }{1+i\omega \tau_\downarrow} , 
\end{eqnarray}
where depending on the sign ($\pm$), the factor $F_{\pm}(\omega)$ represents phenomena related to the charge ($+$) or spin degree of freedom ($-$). For instance, $F_+(\omega)$ determines the total amount of hot electrons that decay within distance $d$ and appears in the description for the local heating process (Sec.\ \ref{sec:5-3}). Furthermore, $F_-(\omega)$ will determine the contribution to the hot electron spin current resulting from the spin-dependent decay rates (Sec.\ \ref{sec:5-4}).  We now have all ingredients to calculate the distinct contributions to the spin current, and to investigate the response of the thermal system to the hot electron dynamics.

\section{Spin pumping mediated by local heating} 
\label{sec:5-3}

In this section, we calculate the spin current that arises in the thermal electron bath and we express it in terms of the characteristic functions for the hot electron dynamics. The method can be separated into three steps. (i) The interfacial spin current within the thermal electron system is expressed in terms of the electron-magnon scattering rate. (ii) The scattering rate is parametrized by an  electron temperature and a magnon temperature. (iii) The magnon temperature is eliminated and the electron temperature is expressed in terms of the hot electron functions defined in the previous section (such as $F_+(\omega)$ from Eq.\ (\ref{eq:5-6})). Combining these three steps yields a simple expression for the thermal spin current that can directly be compared to the hot electron contribution. 

Hence, the starting point is to express the interfacial spin current in terms of the bulk electron-magnon scattering rate. To find a simple description, we assume that the ferromagnetic system is much thinner than the spin-diffusion length, having $d\ll \lambda_{\mathrm{sf}}$.\footnote{It should be noted that we keep the spin-flip scattering rate $\tau_s$ fixed, meaning that the limit $d\ll\lambda_\mathbf{sf}$ actually corresponds to assuming a very large conductivity. The expressions presented here are equivalent to the similar calculation in Ref.\ \cite{Beens2022} for $d\ll \lambda_\mathrm{sf}$. } Then by approximation, the spin density in the thermal electron system is parametrized by a spatial homogeneous spin accumulation $\mu_s$. From the conservation of spin in the combined system, we write down the equation for the out-of-equilibrium spin density $\delta n_{t,s}$ of the thermalized electrons. In the frequency domain it is given by \cite{Tveten2015,Beens2022}
\begin{eqnarray}
\label{eq:5-7}
i\omega \delta n_{t,s} (\omega) +
\dfrac{j_{t,s}(\omega)}{d} 
&=&-2 I_{sd}(\omega)   
- 2 P_A A_0(\omega) 
\\
\nonumber 
&& + 
\dfrac{n_{\uparrow}(\omega) }{\tau_{\uparrow}} 
-\dfrac{n_{\downarrow}(\omega) }{\tau_{\downarrow}}
-\dfrac{\delta n_{t,s}(\omega)}{\tau_s},
\end{eqnarray}
where $j_{t,s}(\omega)$ is the interfacial spin current generated in the thermal electron system. On the right-hand side, $I_{sd}$ is determined by the rate of spin transfer per unit volume driven by electron-magnon scattering \cite{Tveten2015,Bender2012}. The term proportional to $P_A$ corresponds to the spin-dependent excitation of electrons which are transferred to the hot electron system.  Moreover, the terms proportional to $\tau^{-1}_{\uparrow\downarrow}$ result from the decay of the hot electrons. In combination with the previous term (proportional to $P_A$), the latter will generate an additional spin current that will partially compensate the hot electron contribution. This `backflow' spin current will be discussed below. Finally, the last term on the right-hand side of Eq.\ (\ref{eq:5-7}) corresponds to the additional channels of spin-flip scattering with corresponding timescale $\tau_s$ \cite{Tveten2015}. 

The out-of-equilibrium spin density is proportional to the spin accumulation $\delta n_{t,s} = \tilde{\nu}_F \mu_s $, where $ \tilde{\nu}_F$ is the spin-averaged density of states evaluated at the Fermi energy. Analogously, the interfacial spin current carried by the thermal electrons is written as $j_{t,s}(\omega)=(g/\hbar) \mu_s(\omega) $, where $g$ is a conductance for the spin current.  Here, it is assumed that the neighboring nonmagnetic material is a good spin sink. By solving Eq.\ (\ref{eq:5-7}) for $\mu_s$, and using the expressions for $n_\uparrow(\omega)$ and $n_\downarrow(\omega)$ as given in Eq.\ (\ref{eq:5-5}), the interfacial spin current becomes
\begin{equation}
\label{eq:5-8}
j_{t,s}(\omega) = 
\dfrac{(-2 d) I_{sd} }{1+\dfrac{\tau_g}{\tau_s}(1+i\omega \tau_s) } 
 - \dfrac{2P_A A_0(\omega) + A_0(\omega) F_{-}(\omega)  }{1+\dfrac{\tau_g}{\tau_s}(1+i\omega \tau_s)},
\end{equation}
where the timescale $\tau_g $ is defined as $\tau_g^{-1} = g/(\hbar \tilde{\nu}_F d)$ and determines the efficiency of the spin transfer into the nonmagnetic layer. This timescale is treated as an effective parameter to compensate for the fact that (diffusive) spin transport in the bulk is assumed to be instantaneous, as a result of the condition $d\ll \lambda_\mathrm{sf}$.  

\begin{figure}[t!]
\includegraphics[scale=0.95]{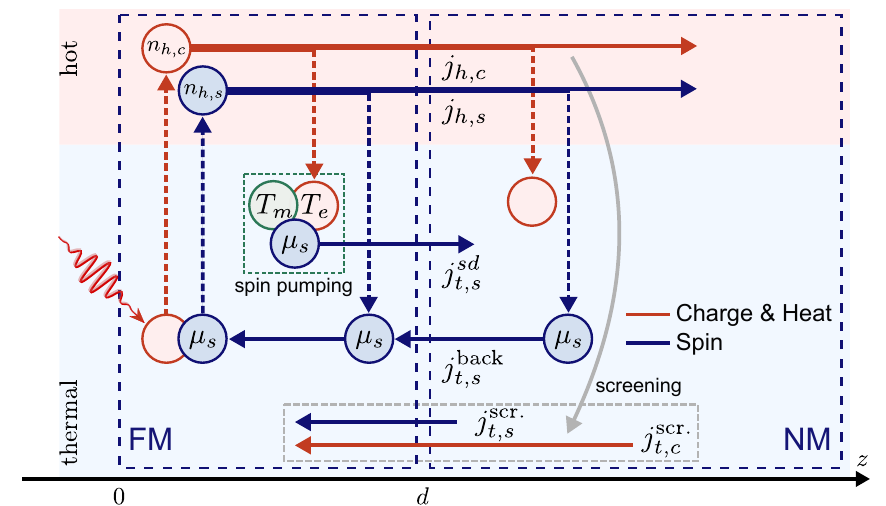}
\caption{\label{fig:5-5} Schematic overview of the various contributions to the laser-induced spin current. The horizontal axis indicates the position within a ferromagnetic/nonmagnetic heterostructure with the interface located at $z=d$. The red and blue shaded regions indicate the local and nonlocal behavior of the `hot' and `thermal' electrons, respectively. Initially, (spin-polarized) hot electrons are excited near the interface at $z=0$, resulting in a charge current $j_{h,c}$ (red solid arrow) and a spin current $j_{h,s}$ (blue solid arrow). Hot electron decay results in the local transfer of heat (red dashed arrows) and spin  (blue dashed arrows) to the thermal system. The former drives a spin current $j_{t,s}^\mathrm{sd}$ through spin pumping (the green box), whereas the latter contributes to the spin current $j_{t,s}^\mathrm{back}$. Additionally, a spin-polarized screening current $j_{t,s}^\mathrm{scr.}$ (the gray box) arises in response to hot electron charge transport.  
} 
\end{figure}

The first term in Eq.\ (\ref{eq:5-8}) corresponds to the spin current driven by the electron-magnon scattering in the bulk (spin pumping) \cite{Tveten2015,Shin2018}, and indirectly results from the local heating process. In Fig.\ \ref{fig:5-5} this contribution is denoted as $j_{t,s}^\mathrm{sd}$. The second term in Eq.\ (\ref{eq:5-8}) is generated because the spin-dependent excitation and decay of hot electrons affect the net spin density in the thermal system, and corresponds to the previously mentioned backflow spin current. In Fig.\ \ref{fig:5-5} it is denoted as $j_{t,s}^\mathrm{back}$. Although the latter is directly related to the hot electron dynamics, it should still be considered  as a spin current contribution carried by thermal electrons. 

To get an analytical expression for the thermal spin current in terms of the excitation profile $A_0(\omega)$, we have to find a simplified description for the electron-magnon scattering rate.  
In order to do so, we first calculate the dynamics of the electron temperature. In the frequency domain the change of the  (spatially-averaged) electron temperature $\delta T_e(\omega)$ satisfies an equation of the form
\begin{equation}
\label{eq:5-9}
i\omega \delta T_e(\omega) =
\dfrac{\Delta E}{C_e} 
\bigg(
\dfrac{n_\uparrow(\omega)}{\tau_\uparrow} 
+\dfrac{n_\downarrow(\omega)}{\tau_\downarrow} 
\bigg)
-\dfrac{\delta T_e(\omega)}{\tau_e },
\end{equation}
where the factor proportional to $\tau_e^{-1}$ is introduced phenomenologically and includes all processes that drive heat out of the electron system in the ferromagnetic region (including heat lost at the interface). Furthermore, $C_e$ is the electronic specific heat and $\Delta E$ is the photon energy of the laser pulse. It follows that the elevated electron temperature $\delta T_e(\omega)$ can be expressed as 
\begin{eqnarray}
\label{eq:5-10} 
\delta T_e(\omega) 
&=& 
\dfrac{\tau_e A_0(\omega)\Delta E }{ d C_e(1+i\omega \tau_e) }
 F_{+}(\omega) .
\end{eqnarray}
To calculate the spin current that results from the increase of the electron temperature, we have to determine the electron-magnon scattering rate $I_{sd}$. We take a simplified approach and assume that the density of magnons that is generated is given by $\delta n_d(\omega) = C_{n,T} \delta T_m(\omega) $, where $\delta T_m(\omega)$ is the Fourier transform of the change of the magnon temperature and the coefficient $C_{n,T}$ is given in Ref.\ \cite{Beens2022} . The rate at which magnons are generated is given by 
\begin{eqnarray}
\label{eq:5-11}
i\omega C_n \delta T_m(\omega) &=& I_{sd}(\omega) 
= \dfrac{C_{n,T} }{\tau_m} 
(\delta T_e(\omega)-\delta T_m(\omega)), 
\end{eqnarray}
where the electron-magnon scattering rate is expressed in terms of the difference in magnon temperature and electron temperature, and is proportional to a corresponding (demagnetization) timescale $\tau_m$. Combining Eq.\ (\ref{eq:5-10}) and Eq.\ (\ref{eq:5-11}) gives a closed expression for the electron-magnon scattering rate in terms of the functions that depend on the hot electron system. This yields
\begin{eqnarray}
\label{eq:5-12}
I_{sd}(\omega) &=& \dfrac{C_{n,T}\Delta E}{C_e } 
\dfrac{(i\omega\tau_e) A_0(\omega)   F_+(\omega) }{d (1+i\omega\tau_e)(1+i\omega\tau_m)}.
\end{eqnarray}
Physically, the product describes the consecutive processes of heating the thermal electrons through the energy retrieved from  decaying hot electrons (described by $F^+(\omega)$), and the subsequent generation of thermal magnons by an increase of the temperature. By substituting the expression for $I_{sd}(\omega)$ in Eq.\ (\ref{eq:5-8}), the spin current driven by electron-magnon scattering can be expressed in terms of the functions that describe the hot electron dynamics.

\section{Comparison of the hot and thermal spin currents} 
\label{sec:5-4}

In this section, we directly compare the distinct contributions  to the interfacial spin current. First, the total interfacial spin current is written as 
\begin{equation}
\label{eq:5-13} 
j_s^\mathrm{tot}(\omega) = j_{h,s}(\omega)+ j_{t,s}^\mathrm{sd} (\omega) 
+  j_{t,s}^\mathrm{back} (\omega),
\end{equation}
where $j_{h,s}$ corresponds to the direct spin current carried by hot electrons and $j_{t,s}^\mathrm{sd}$ corresponds to the thermal contribution driven by spin pumping. The spin current $j_{t,s}^\mathrm{back}$ is equal to the second term on the right-hand side of Eq.\ (\ref{eq:5-8}), named after that it drives a backflow that partially compensates the hot electron contribution. All the given contributions to the interfacial spin current are represented in the schematic overview in Fig.\ \ref{fig:5-5}.

\begin{figure}[t!]
\includegraphics[scale=1.00]{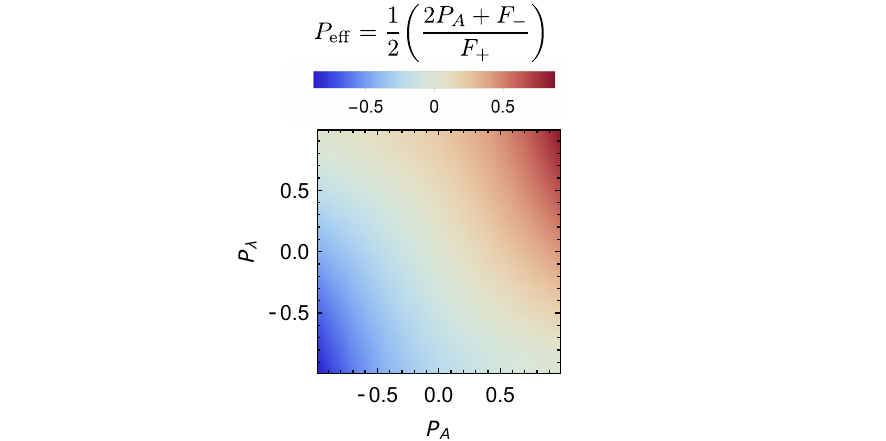}
\caption{\label{fig:5-2-b} The effective polarization function $P_\mathrm{eff}$ (discussed below Eq.\ (\ref{eq:5-18})) as a function of the polarization of the decay lengths $P_\lambda$ and the polarization of the excitation $P_A$.  
} 
\end{figure}

\begin{figure*}[t]
\includegraphics[scale=0.95]{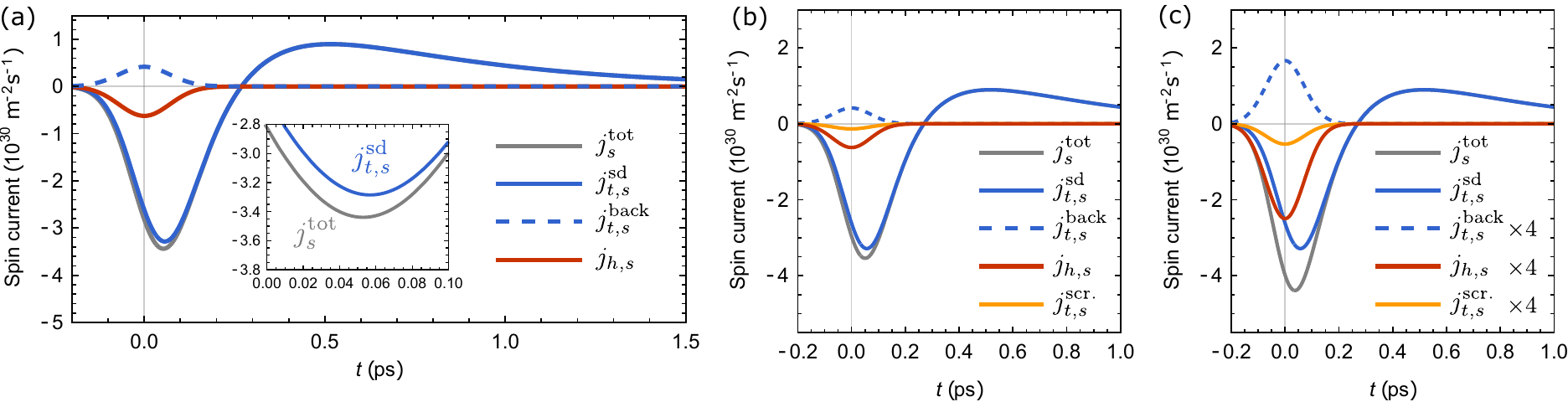}
\caption{\label{fig:5-3} The interfacial spin current as a function of time after laser-pulse excitation. The gray line indicates the total spin current. Furthermore, the  blue line indicates the contribution driven by  electron-magnon scattering and the red line shows the hot electron spin current. Finally, the dashed blue line corresponds to the backflow current, as defined in the main text. (b) Similar calculation as (a), but now including a spin-polarized screening current indicated by the yellow line. (c) The contributions (indirectly) related to hot electron dynamics are multiplied by a factor of four to visually clarify the role of each contribution and to show the change of the total spin current (gray).  
} 
\end{figure*}

To derive a simple relation that parametrizes the ratio between the different contributions to the spin current, we make the following assumptions. First, we assume that the ferromagnetic layer is very thin such that $\tau_g $ satisfies $\tau_g \omega\ll 1$. Similarly, we assume that the decay rate of the hot electrons is very fast $\tau_{\uparrow\downarrow}\omega \ll 1$. This means that we model a laser pulse that has a duration $\sigma\gg \tau_g,\tau_{\uparrow\downarrow} $. In that scenario we find
\begin{eqnarray}
\label{eq:5-14}
j_{h,s}(\omega) &=& 2 P_A A_0(\omega)+ A_0(\omega) 
F_-(0) ,
\\
\label{eq:5-15}
j_{t,s}^\mathrm{sd} (\omega) &=& 
\dfrac{-2 C_{n,T}\Delta E}{C_e\big(1+\tau_g/\tau_s\big) } 
\dfrac{(i\omega\tau_e ) A_0(\omega)   F_+(0) }{  (1+i\omega\tau_e)(1+i\omega\tau_m) },
\\
\label{eq:5-16}
j_{t,s}^\mathrm{back}(\omega) &=& -
\dfrac{2 P_A A_0(\omega) + A_0(\omega)F_{-}(0) }{1+\tau_g/\tau_s} .
\end{eqnarray}
Importantly, it shows that $j_{t,s}^\mathrm{back}$ is directly proportional to the hot electron contribution and has an opposite sign. To explicitly calculate the spin currents in the time domain we assume the following temporal profile of the laser pulse
\begin{eqnarray}
\label{eq:5-17}
A_0(t) &=& \dfrac{P_0 d}{\Delta E (\sigma \sqrt{\pi})} 
\exp(-t^2/\sigma^2 ),
\end{eqnarray} 
where $\sigma$ is the pulse duration, $P_0$ plays the role of an absorbed laser pulse energy density, and $\Delta E$ is the photon energy. Inverse Fourier transforming Eqs.\ (\ref{eq:5-14})-(\ref{eq:5-16}) (and for Eq.\ (\ref{eq:5-15}) performing a convolution in the time domain) yields the temporal profiles of the distinct spin current contributions. Figure \ref{fig:5-3}(a)  shows the resulting interfacial spin current as a function of time after laser-pulse excitation at $t=0$. The used system parameters are presented in Table  \ref{tab:5-1}, which represent a typical magnetic heterostructure consisting of transition metal ferromagnet and a  nonmagnetic metal that is a good spin sink (such as Pt). In the figure, the gray line indicates the total spin current and the blue line shows the contribution by spin pumping. Furthermore, the red line represents the hot electron spin current and the dashed blue line the backflow spin current. The figure shows that for the used parameters the total spin current is dominated by the spin pumping contribution. The amplitude of the latter is approximately a factor $\sim 5$ times larger than the hot electron contribution. Moreover, including the backflow spin current yields that the hot electron spin current is almost completely compensated. 

To further investigate the role of the several contributions to spin transport, it is convenient to calculate the ratio between $j_{t,s}^\mathrm{sd}$ and $j_{h,s}$ from Eq.\ (\ref{eq:5-14}) and Eq.\ (\ref{eq:5-15}). We define $\eta$ as   
\begin{equation}
\label{eq:5-18}
\dfrac{\mbox{max}(|j_{t,s}^\mathrm{sd}|)}{\mbox{max}(|j_{h,s}|)} 
 \equiv 
  \eta  \propto   \dfrac{C_{n,T} \Delta E }{C_e\Big(1+\tau_g/\tau_s \Big)} 
\bigg[\dfrac{-F_+(0)}{2 P_A + F_{-}(0)} \bigg].
\end{equation}
Note that the exact ratio of the amplitudes also includes an additional prefactor determined by $\sigma$, $\tau_e$ and $\tau_m$ (not included in Eq.\ (\ref{eq:5-18})). As is shown in Appendix \ref{sec:app5-A}, this additional factor typically scales as $\sigma/\tau_m$, which in our example is of the order of one. The term between square brackets, on the right-hand side of Eq.\ (\ref{eq:5-18}),  plays the role of an effective polarization $P_{\mathrm{eff}}$ of the hot electron current, and is determined by $P_A$ and $P_\lambda=(\lambda_\uparrow-\lambda_\downarrow)/(\lambda_\uparrow+\lambda_\downarrow)$. This $P_\mathrm{eff}$ is plotted in Fig.\ \ref{fig:5-2-b} as a function of $P_A$ and $P_\lambda$,  for $d/\lambda=0.3$ with $\lambda=(\lambda_\uparrow+\lambda_\downarrow)/2$. $P_\mathrm{eff}$ is shown to be a monotonic function of $P_A$ and $P_\lambda$, which explains why it is interpreted as an effective polarization.

To express the ratio $\eta$ in terms of fundamental parameters,  we use that for a free electron gas the specific heat scales as  $C_e \sim  k_B (T/T_F)/a^3 $ \cite{Ashcroft1976}, where $a$ is the lattice constant and $T_F$ is the Fermi temperature. Furthermore, the magnon density coefficient $C_{n,T}$ scales as $C_{n,T}\sim (k_B T)^{1/2} A^{-3/2}k_B $ \cite{Beens2022}, where the spin-wave stiffness is proportional to the Curie temperature $A\sim k_B T_C a^2$. By implementing the numerical prefactors (including multiple factors of $\pi$) we estimate the order of magnitude of $\eta$ and determine the crucial scaling factors 

\begin{equation}
\label{eq:5-19} 
\eta  \approx 
\Bigg[ 
\dfrac{2*10^{-2}}{(-P_\mathrm{eff}) } 
\dfrac{1}
{1+\tau_g/\tau_s }
\Bigg] 
\bigg(\dfrac{\Delta E}{k_B T} \bigg)
\bigg(\dfrac{T_F}{T_C} \bigg)
\sqrt{\dfrac{T}{T_C}} . 
\end{equation}

Although the factor between square brackets yields a number much smaller than one, this number is largely compensated by the remaining factors. Specifically, for a transition metal ferromagnet  the Fermi temperature and Curie temperature typically differ an order of magnitude $(T_F/T_C)\sim 10$. Furthermore, for a $\Delta E$ of the order of electronvolts and a temperature close to room temperature we find the range $\Delta E/(k_B T)\sim 10^2\mbox{-}10^3$.  Altogether, this implies that the contribution by spin pumping is generally large compared to the contribution by hot electron transport. In case the backflow is taken into account, the partial compensation of the hot electron spin current would lead to a change in the prefactor $(1+\tau_g/\tau_s)^{-1}\rightarrow \tau_s/\tau_g $, resulting in an even  larger $\eta$ for $\tau_s>\tau_g$.


\section{The role of spin-polarized screening currents} 

Finally, we discuss the role of spin-polarized screening. It is generally assumed that screening of the charge degree of freedom happens on an extremely short timescale \cite{Kimling2017}. This corresponds to the approximation in the model that the system remains locally charge neutral and that the total charge current of the hot and thermal electrons is zero at all times. In the case of charge transport in the thermal electron system, the efficient screening approximation was already implemented throughout the previous sections. Additionally, in this work, we have the excited hot electrons that carry a nonzero charge current for which we analogously assume it is effectively screened through transport in the thermal electron system. This process is schematically depicted in the gray box in Fig.\ \ref{fig:5-5}. Within the ferromagnetic region it results in an extra contribution to the spin current since the present screening currents are subject to spin-dependent transport coefficients. Implementing spin-polarized screening currents within the toy model yields the following extension. First, charge neutrality requires the spin density of the thermalized system to satisfy 
\begin{eqnarray}
\label{eq:5-20}
\delta n_s &=& \tilde{\nu}_F \mu_s 
-P_\nu (n_\uparrow+n_\downarrow),
\end{eqnarray}
where $P_\nu=(\nu_\uparrow-\nu_\downarrow)/(\nu_\uparrow+\nu_\downarrow) $ corresponds to the polarization of the density of states at the Fermi energy. Secondly, the absence of a net charge current requires that 
\begin{eqnarray}
\label{eq:5-21} 
j_{t,s} &=& \dfrac{g}{\hbar } 
\mu_s - P_g (j_\uparrow+j_\downarrow) ,
\end{eqnarray}
where we defined $P_g =(g_\uparrow-g_\downarrow)/(g_\uparrow+g_\downarrow)$. The  conductance we used previously is given by the spin-averaged conductance $g=2 g_\uparrow g_\downarrow/(g_\uparrow+g_\downarrow)$. Implementing this within the previous scheme for the thermal electron system yields an extra contribution to the spin current 
\begin{eqnarray}
\label{eq:5-22}
j_{s,t}^\mathrm{scr.}(\omega) &=& -P_g (j_\uparrow(\omega)+j_\downarrow(\omega) )
\dfrac{(\tau_g/\tau_s) }{1+(\tau_g/\tau_s)  }   .
\end{eqnarray}
This is the spin-polarized screening current. Here, the term proportional to $P_\nu$ vanished due to the limit $\omega\tau_{\uparrow\downarrow}\ll 1$. The spin-polarized screening current is calculated for $P_g=0.2$ and represented by the yellow curve in Fig.\ \ref{fig:5-3}(b). Depending on the sign of $P_g$ this contribution to the spin current either enhances or partially compensates the hot electron contribution. For illustrative purposes, Fig.\ \ref{fig:5-3}(b)  includes all other contributions to the spin current.

Additionally, we included Fig.\ \ref{fig:5-3}(c). Here, we multiplied all terms (indirectly) related to the hot electron dynamics by a factor of four to emphasize the role of each separate contribution and to show the change of the total spin current (in gray). The figure emphasizes that in the case that the hot electron spin current is enhanced, for instance when taking into account multiple generations of hot electrons, the total spin current is significantly modified. Nevertheless, for the parameters used here, the spin current driven by spin pumping remains the dominant contribution. 

\section{Conclusion and outlook} 

In conclusion, using a single simplified analytical model, we investigated the role of spin-polarized hot electron transport and spin transport driven by spin pumping in laser-excited magnetic heterostructures. This toy model yields that the spin current at the interface of the heterostructure is dominated by the thermal contribution initiated by local heating and subsequent spin pumping. We calculated the scaling factors that determine the ratio between the two contributions. As the latter depends on the fundamental parameters that describe the magnon system and thermal electron system, it could be expressed in terms of the Curie temperature, Fermi temperature, and laser-photon energy. This fundamental relation yields that the spin current driven by spin pumping is generally a significant contribution, and is dominant for the systems considered here.    

An interesting extension to the toy model would be to implement multiple generations of hot electrons and calculate the resulting enhancement of the spin current. In that way, one reaches a description similar to the model for superdiffusive spin transport \cite{Battiato2010,Battiato2012}. Additionally, it would be interesting to implement the conceptual spin-polarized screening currents within the superdiffusive approach. Moreover, spin transport by thermal magnons and interfacial electron-magnon scattering processes are required to be investigated within this scheme \cite{Tveten2015,Beens2022}. Nevertheless, it is expected that those extensions leave the presented scaling factors intact and spin pumping through local heating remains a dominant channel for spin-current generation.       

\section{Acknowledgments}

This work is part of the research programme of the Foundation for Fundamental Research on Matter (FOM), which is part of the Netherlands Organisation for Scientific Research (NWO). R.D. is member of the D-ITP consortium, a program of the NWO that is funded by the Dutch Ministry of Education, Culture and Science (OCW). This work is funded by the European Research Council (ERC).

\bibliographystyle{apsrev4-2}

\begin{thebibliography}{44}%
\makeatletter
\providecommand \@ifxundefined [1]{%
 \@ifx{#1\undefined}
}%
\providecommand \@ifnum [1]{%
 \ifnum #1\expandafter \@firstoftwo
 \else \expandafter \@secondoftwo
 \fi
}%
\providecommand \@ifx [1]{%
 \ifx #1\expandafter \@firstoftwo
 \else \expandafter \@secondoftwo
 \fi
}%
\providecommand \natexlab [1]{#1}%
\providecommand \enquote  [1]{``#1''}%
\providecommand \bibnamefont  [1]{#1}%
\providecommand \bibfnamefont [1]{#1}%
\providecommand \citenamefont [1]{#1}%
\providecommand \href@noop [0]{\@secondoftwo}%
\providecommand \href [0]{\begingroup \@sanitize@url \@href}%
\providecommand \@href[1]{\@@startlink{#1}\@@href}%
\providecommand \@@href[1]{\endgroup#1\@@endlink}%
\providecommand \@sanitize@url [0]{\catcode `\\12\catcode `\$12\catcode
  `\&12\catcode `\#12\catcode `\^12\catcode `\_12\catcode `\%12\relax}%
\providecommand \@@startlink[1]{}%
\providecommand \@@endlink[0]{}%
\providecommand \url  [0]{\begingroup\@sanitize@url \@url }%
\providecommand \@url [1]{\endgroup\@href {#1}{\urlprefix }}%
\providecommand \urlprefix  [0]{URL }%
\providecommand \Eprint [0]{\href }%
\providecommand \doibase [0]{https://doi.org/}%
\providecommand \selectlanguage [0]{\@gobble}%
\providecommand \bibinfo  [0]{\@secondoftwo}%
\providecommand \bibfield  [0]{\@secondoftwo}%
\providecommand \translation [1]{[#1]}%
\providecommand \BibitemOpen [0]{}%
\providecommand \bibitemStop [0]{}%
\providecommand \bibitemNoStop [0]{.\EOS\space}%
\providecommand \EOS [0]{\spacefactor3000\relax}%
\providecommand \BibitemShut  [1]{\csname bibitem#1\endcsname}%
\let\auto@bib@innerbib\@empty
\bibitem [{\citenamefont {Beaurepaire}\ \emph {et~al.}(1996)\citenamefont
  {Beaurepaire}, \citenamefont {{J.-C. Merle}}, \citenamefont {Daunois},\ and\
  \citenamefont {{J.-Y. Bigot}}}]{Beaurepaire1996}%
  \BibitemOpen
  \bibfield  {author} {\bibinfo {author} {\bibfnamefont {E.}~\bibnamefont
  {Beaurepaire}}, \bibinfo {author} {\bibnamefont {{J.-C. Merle}}}, \bibinfo
  {author} {\bibfnamefont {A.}~\bibnamefont {Daunois}},\ and\ \bibinfo {author}
  {\bibnamefont {{J.-Y. Bigot}}},\ }\href
  {https://doi.org/10.1103/PhysRevLett.76.4250} {\bibfield  {journal} {\bibinfo
   {journal} {Phys. Rev. Lett.}\ }\textbf {\bibinfo {volume} {76}},\ \bibinfo
  {pages} {4250} (\bibinfo {year} {1996})}\BibitemShut {NoStop}%
\bibitem [{\citenamefont {{G.P. Zhang}}\ and\ \citenamefont
  {H{\"u}bner}(2000)}]{Zhang2000}%
  \BibitemOpen
  \bibfield  {author} {\bibinfo {author} {\bibnamefont {{G.P. Zhang}}}\ and\
  \bibinfo {author} {\bibfnamefont {W.}~\bibnamefont {H{\"u}bner}},\ }\href
  {https://doi.org/10.1103/PhysRevLett.85.3025} {\bibfield  {journal} {\bibinfo
   {journal} {Phys. Rev. Lett.}\ }\textbf {\bibinfo {volume} {85}},\ \bibinfo
  {pages} {3025} (\bibinfo {year} {2000})}\BibitemShut {NoStop}%
\bibitem [{\citenamefont {Koopmans}\ \emph {et~al.}(2005)\citenamefont
  {Koopmans}, \citenamefont {{J.J.M. Ruigrok}}, \citenamefont {{Dalla Longa}},\
  and\ \citenamefont {{W.J.M. de Jonge}}}]{Koopmans2005}%
  \BibitemOpen
  \bibfield  {author} {\bibinfo {author} {\bibfnamefont {B.}~\bibnamefont
  {Koopmans}}, \bibinfo {author} {\bibnamefont {{J.J.M. Ruigrok}}}, \bibinfo
  {author} {\bibfnamefont {F.}~\bibnamefont {{Dalla Longa}}},\ and\ \bibinfo
  {author} {\bibnamefont {{W.J.M. de Jonge}}},\ }\href
  {https://doi.org/10.1103/PhysRevLett.95.267207} {\bibfield  {journal}
  {\bibinfo  {journal} {Phys. Rev. Lett.}\ }\textbf {\bibinfo {volume} {95}},\
  \bibinfo {pages} {267207} (\bibinfo {year} {2005})}\BibitemShut {NoStop}%
\bibitem [{\citenamefont {Kazantseva}\ \emph {et~al.}(2007)\citenamefont
  {Kazantseva}, \citenamefont {Nowak}, \citenamefont {{R.W. Chantrell}},
  \citenamefont {Hohlfeld},\ and\ \citenamefont {Rebei}}]{Kazantseva2007}%
  \BibitemOpen
  \bibfield  {author} {\bibinfo {author} {\bibfnamefont {N.}~\bibnamefont
  {Kazantseva}}, \bibinfo {author} {\bibfnamefont {U.}~\bibnamefont {Nowak}},
  \bibinfo {author} {\bibnamefont {{R.W. Chantrell}}}, \bibinfo {author}
  {\bibfnamefont {J.}~\bibnamefont {Hohlfeld}},\ and\ \bibinfo {author}
  {\bibfnamefont {A.}~\bibnamefont {Rebei}},\ }\href
  {https://doi.org/10.1209/0295-5075/81/27004} {\bibfield  {journal} {\bibinfo
  {journal} {Europhys. Lett.}\ }\textbf {\bibinfo {volume} {81}},\ \bibinfo
  {pages} {27004} (\bibinfo {year} {2007})}\BibitemShut {NoStop}%
\bibitem [{\citenamefont {Carpene}\ \emph {et~al.}(2008)\citenamefont
  {Carpene}, \citenamefont {Mancini}, \citenamefont {Dallera}, \citenamefont
  {Brenna}, \citenamefont {Puppin},\ and\ \citenamefont
  {De~Silvestri}}]{carpene2008dynamics}%
  \BibitemOpen
  \bibfield  {author} {\bibinfo {author} {\bibfnamefont {E.}~\bibnamefont
  {Carpene}}, \bibinfo {author} {\bibfnamefont {E.}~\bibnamefont {Mancini}},
  \bibinfo {author} {\bibfnamefont {C.}~\bibnamefont {Dallera}}, \bibinfo
  {author} {\bibfnamefont {M.}~\bibnamefont {Brenna}}, \bibinfo {author}
  {\bibfnamefont {E.}~\bibnamefont {Puppin}},\ and\ \bibinfo {author}
  {\bibfnamefont {S.}~\bibnamefont {De~Silvestri}},\ }\href
  {https://doi.org/10.1103/PhysRevB.78.174422} {\bibfield  {journal} {\bibinfo
  {journal} {Phys. Rev. B}\ }\textbf {\bibinfo {volume} {78}},\ \bibinfo
  {pages} {174422} (\bibinfo {year} {2008})}\BibitemShut {NoStop}%
\bibitem [{\citenamefont {{J.-Y. Bigot}}\ \emph {et~al.}(2009)\citenamefont
  {{J.-Y. Bigot}}, \citenamefont {Vomir},\ and\ \citenamefont
  {Beaurepaire}}]{Bigot2009}%
  \BibitemOpen
  \bibfield  {author} {\bibinfo {author} {\bibnamefont {{J.-Y. Bigot}}},
  \bibinfo {author} {\bibfnamefont {M.}~\bibnamefont {Vomir}},\ and\ \bibinfo
  {author} {\bibfnamefont {E.}~\bibnamefont {Beaurepaire}},\ }\href
  {https://doi.org/10.1038/nphys1285} {\bibfield  {journal} {\bibinfo
  {journal} {Nat. Phys.}\ }\textbf {\bibinfo {volume} {5}},\ \bibinfo {pages}
  {515} (\bibinfo {year} {2009})}\BibitemShut {NoStop}%
\bibitem [{\citenamefont {Krau{\ss}}\ \emph {et~al.}(2009)\citenamefont
  {Krau{\ss}}, \citenamefont {Roth}, \citenamefont {Alebrand}, \citenamefont
  {Steil}, \citenamefont {Cinchetti}, \citenamefont {Aeschlimann},\ and\
  \citenamefont {{H.C. Schneider}}}]{Krauss2009}%
  \BibitemOpen
  \bibfield  {author} {\bibinfo {author} {\bibfnamefont {M.}~\bibnamefont
  {Krau{\ss}}}, \bibinfo {author} {\bibfnamefont {T.}~\bibnamefont {Roth}},
  \bibinfo {author} {\bibfnamefont {S.}~\bibnamefont {Alebrand}}, \bibinfo
  {author} {\bibfnamefont {D.}~\bibnamefont {Steil}}, \bibinfo {author}
  {\bibfnamefont {M.}~\bibnamefont {Cinchetti}}, \bibinfo {author}
  {\bibfnamefont {M.}~\bibnamefont {Aeschlimann}},\ and\ \bibinfo {author}
  {\bibnamefont {{H.C. Schneider}}},\ }\href
  {https://doi.org/10.1103/PhysRevB.80.180407} {\bibfield  {journal} {\bibinfo
  {journal} {Phys. Rev. B}\ }\textbf {\bibinfo {volume} {80}},\ \bibinfo
  {pages} {180407(R)} (\bibinfo {year} {2009})}\BibitemShut {NoStop}%
\bibitem [{\citenamefont {Koopmans}\ \emph {et~al.}(2010)\citenamefont
  {Koopmans}, \citenamefont {Malinowski}, \citenamefont {{Dalla Longa}},
  \citenamefont {Steiauf}, \citenamefont {F{\"a}hnle}, \citenamefont {Roth},
  \citenamefont {Cinchetti},\ and\ \citenamefont {Aeschlimann}}]{Koopmans2010}%
  \BibitemOpen
  \bibfield  {author} {\bibinfo {author} {\bibfnamefont {B.}~\bibnamefont
  {Koopmans}}, \bibinfo {author} {\bibfnamefont {G.}~\bibnamefont
  {Malinowski}}, \bibinfo {author} {\bibfnamefont {F.}~\bibnamefont {{Dalla
  Longa}}}, \bibinfo {author} {\bibfnamefont {D.}~\bibnamefont {Steiauf}},
  \bibinfo {author} {\bibfnamefont {M.}~\bibnamefont {F{\"a}hnle}}, \bibinfo
  {author} {\bibfnamefont {T.}~\bibnamefont {Roth}}, \bibinfo {author}
  {\bibfnamefont {M.}~\bibnamefont {Cinchetti}},\ and\ \bibinfo {author}
  {\bibfnamefont {M.}~\bibnamefont {Aeschlimann}},\ }\href
  {https://doi.org/10.1038/NMAT2593} {\bibfield  {journal} {\bibinfo  {journal}
  {Nat. Mat.}\ }\textbf {\bibinfo {volume} {9}},\ \bibinfo {pages} {259}
  (\bibinfo {year} {2010})}\BibitemShut {NoStop}%
\bibitem [{\citenamefont {Battiato}\ \emph {et~al.}(2010)\citenamefont
  {Battiato}, \citenamefont {Carva},\ and\ \citenamefont
  {Oppeneer}}]{Battiato2010}%
  \BibitemOpen
  \bibfield  {author} {\bibinfo {author} {\bibfnamefont {M.}~\bibnamefont
  {Battiato}}, \bibinfo {author} {\bibfnamefont {K.}~\bibnamefont {Carva}},\
  and\ \bibinfo {author} {\bibfnamefont {P.~M.}\ \bibnamefont {Oppeneer}},\
  }\href {https://doi.org/10.1103/PhysRevLett.105.027203} {\bibfield  {journal}
  {\bibinfo  {journal} {Phys. Rev. Lett.}\ }\textbf {\bibinfo {volume} {105}},\
  \bibinfo {pages} {027203} (\bibinfo {year} {2010})}\BibitemShut {NoStop}%
\bibitem [{\citenamefont {Atxitia}\ and\ \citenamefont
  {Chubykalo-Fesenko}(2011)}]{Atxitia2011}%
  \BibitemOpen
  \bibfield  {author} {\bibinfo {author} {\bibfnamefont {U.}~\bibnamefont
  {Atxitia}}\ and\ \bibinfo {author} {\bibfnamefont {O.}~\bibnamefont
  {Chubykalo-Fesenko}},\ }\href {https://doi.org/10.1103/PhysRevB.84.144414}
  {\bibfield  {journal} {\bibinfo  {journal} {Phys. Rev. B}\ }\textbf {\bibinfo
  {volume} {84}},\ \bibinfo {pages} {144414} (\bibinfo {year}
  {2011})}\BibitemShut {NoStop}%
\bibitem [{\citenamefont {Manchon}\ \emph {et~al.}(2012)\citenamefont
  {Manchon}, \citenamefont {Li}, \citenamefont {Xu},\ and\ \citenamefont
  {Zhang}}]{Manchon2012}%
  \BibitemOpen
  \bibfield  {author} {\bibinfo {author} {\bibfnamefont {A.}~\bibnamefont
  {Manchon}}, \bibinfo {author} {\bibfnamefont {Q.}~\bibnamefont {Li}},
  \bibinfo {author} {\bibfnamefont {L.}~\bibnamefont {Xu}},\ and\ \bibinfo
  {author} {\bibfnamefont {S.}~\bibnamefont {Zhang}},\ }\href
  {https://doi.org/10.1103/PhysRevB.85.064408} {\bibfield  {journal} {\bibinfo
  {journal} {Phys. Rev. B}\ }\textbf {\bibinfo {volume} {85}},\ \bibinfo
  {pages} {064408} (\bibinfo {year} {2012})}\BibitemShut {NoStop}%
\bibitem [{\citenamefont {{B.Y. Mueller}}\ \emph {et~al.}(2013)\citenamefont
  {{B.Y. Mueller}}, \citenamefont {Baral}, \citenamefont {Vollmar},
  \citenamefont {Cinchetti}, \citenamefont {Aeschlimann}, \citenamefont {{H.C.
  Schneider}},\ and\ \citenamefont {Rethfeld}}]{mueller2013feedback}%
  \BibitemOpen
  \bibfield  {author} {\bibinfo {author} {\bibnamefont {{B.Y. Mueller}}},
  \bibinfo {author} {\bibfnamefont {A.}~\bibnamefont {Baral}}, \bibinfo
  {author} {\bibfnamefont {S.}~\bibnamefont {Vollmar}}, \bibinfo {author}
  {\bibfnamefont {M.}~\bibnamefont {Cinchetti}}, \bibinfo {author}
  {\bibfnamefont {M.}~\bibnamefont {Aeschlimann}}, \bibinfo {author}
  {\bibnamefont {{H.C. Schneider}}},\ and\ \bibinfo {author} {\bibfnamefont
  {B.}~\bibnamefont {Rethfeld}},\ }\href
  {https://doi.org/10.1103/PhysRevLett.111.167204} {\bibfield  {journal}
  {\bibinfo  {journal} {Phys. Rev. Lett.}\ }\textbf {\bibinfo {volume} {111}},\
  \bibinfo {pages} {167204} (\bibinfo {year} {2013})}\BibitemShut {NoStop}%
\bibitem [{\citenamefont {{B.Y. Mueller}}\ and\ \citenamefont
  {Rethfeld}(2014)}]{Mueller2014}%
  \BibitemOpen
  \bibfield  {author} {\bibinfo {author} {\bibnamefont {{B.Y. Mueller}}}\ and\
  \bibinfo {author} {\bibfnamefont {B.}~\bibnamefont {Rethfeld}},\ }\href
  {https://doi.org/10.1103/PhysRevB.90.144420} {\bibfield  {journal} {\bibinfo
  {journal} {Phys. Rev. B}\ }\textbf {\bibinfo {volume} {90}},\ \bibinfo
  {pages} {144420} (\bibinfo {year} {2014})}\BibitemShut {NoStop}%
\bibitem [{\citenamefont {Haag}\ \emph {et~al.}(2014)\citenamefont {Haag},
  \citenamefont {Illg},\ and\ \citenamefont {F{\"a}hnle}}]{Haag2014}%
  \BibitemOpen
  \bibfield  {author} {\bibinfo {author} {\bibfnamefont {M.}~\bibnamefont
  {Haag}}, \bibinfo {author} {\bibfnamefont {C.}~\bibnamefont {Illg}},\ and\
  \bibinfo {author} {\bibfnamefont {M.}~\bibnamefont {F{\"a}hnle}},\ }\href
  {https://doi.org/10.1103/PhysRevB.90.014417} {\bibfield  {journal} {\bibinfo
  {journal} {Phys. Rev. B}\ }\textbf {\bibinfo {volume} {90}},\ \bibinfo
  {pages} {014417} (\bibinfo {year} {2014})}\BibitemShut {NoStop}%
\bibitem [{\citenamefont {Nieves}\ \emph {et~al.}(2014)\citenamefont {Nieves},
  \citenamefont {Serantes}, \citenamefont {Atxitia},\ and\ \citenamefont
  {Chubykalo-Fesenko}}]{Nieves2014}%
  \BibitemOpen
  \bibfield  {author} {\bibinfo {author} {\bibfnamefont {P.}~\bibnamefont
  {Nieves}}, \bibinfo {author} {\bibfnamefont {D.}~\bibnamefont {Serantes}},
  \bibinfo {author} {\bibfnamefont {U.}~\bibnamefont {Atxitia}},\ and\ \bibinfo
  {author} {\bibfnamefont {O.}~\bibnamefont {Chubykalo-Fesenko}},\ }\href
  {https://doi.org/10.1103/PhysRevB.90.104428} {\bibfield  {journal} {\bibinfo
  {journal} {Phys. Rev. B}\ }\textbf {\bibinfo {volume} {90}},\ \bibinfo
  {pages} {104428} (\bibinfo {year} {2014})}\BibitemShut {NoStop}%
\bibitem [{\citenamefont {{E.G. Tveten}}\ \emph {et~al.}(2015)\citenamefont
  {{E.G. Tveten}}, \citenamefont {Brataas},\ and\ \citenamefont
  {Tserkovnyak}}]{Tveten2015}%
  \BibitemOpen
  \bibfield  {author} {\bibinfo {author} {\bibnamefont {{E.G. Tveten}}},
  \bibinfo {author} {\bibfnamefont {A.}~\bibnamefont {Brataas}},\ and\ \bibinfo
  {author} {\bibfnamefont {Y.}~\bibnamefont {Tserkovnyak}},\ }\href
  {https://doi.org/10.1103/PhysRevB.92.180412} {\bibfield  {journal} {\bibinfo
  {journal} {Phys. Rev. B}\ }\textbf {\bibinfo {volume} {92}},\ \bibinfo
  {pages} {180412(R)} (\bibinfo {year} {2015})}\BibitemShut {NoStop}%
\bibitem [{\citenamefont {Krieger}\ \emph {et~al.}(2015)\citenamefont
  {Krieger}, \citenamefont {{J.K. Dewhurst}}, \citenamefont {Elliott},
  \citenamefont {Sharma},\ and\ \citenamefont {{E.K.U. Gross}}}]{Krieger2015}%
  \BibitemOpen
  \bibfield  {author} {\bibinfo {author} {\bibfnamefont {K.}~\bibnamefont
  {Krieger}}, \bibinfo {author} {\bibnamefont {{J.K. Dewhurst}}}, \bibinfo
  {author} {\bibfnamefont {P.}~\bibnamefont {Elliott}}, \bibinfo {author}
  {\bibfnamefont {S.}~\bibnamefont {Sharma}},\ and\ \bibinfo {author}
  {\bibnamefont {{E.K.U. Gross}}},\ }\href
  {https://doi.org/10.1021/acs.jctc.5b00621} {\bibfield  {journal} {\bibinfo
  {journal} {Journal of chemical theory and computation}\ }\textbf {\bibinfo
  {volume} {11}},\ \bibinfo {pages} {4870} (\bibinfo {year}
  {2015})}\BibitemShut {NoStop}%
\bibitem [{\citenamefont {T{\"o}ws}\ and\ \citenamefont {{G.M.
  Pastor}}(2015)}]{Tows2015}%
  \BibitemOpen
  \bibfield  {author} {\bibinfo {author} {\bibfnamefont {W.}~\bibnamefont
  {T{\"o}ws}}\ and\ \bibinfo {author} {\bibnamefont {{G.M. Pastor}}},\ }\href
  {https://doi.org/10.1103/PhysRevLett.115.217204} {\bibfield  {journal}
  {\bibinfo  {journal} {Phys. Rev. Lett.}\ }\textbf {\bibinfo {volume} {115}},\
  \bibinfo {pages} {217204} (\bibinfo {year} {2015})}\BibitemShut {NoStop}%
\bibitem [{\citenamefont {Dornes}\ \emph {et~al.}(2019)\citenamefont {Dornes},
  \citenamefont {Acremann}, \citenamefont {Savoini}, \citenamefont {Kubli},
  \citenamefont {{M.J. Neugebauer}}, \citenamefont {Abreu}, \citenamefont
  {Huber}, \citenamefont {Lantz}, \citenamefont {{C.A.F. Vaz}}, \citenamefont
  {Lemke}, \citenamefont {{E.M. Bothschafter}}, \citenamefont {{M. Porer}},
  \citenamefont {{V. Esposito}}, \citenamefont {{L. Rettig}}, \citenamefont
  {{M. Buzzi}}, \citenamefont {{A. Alberca}}, \citenamefont {{Y.W. Windsor}},
  \citenamefont {{P. Beaud}}, \citenamefont {{U. Staub}}, \citenamefont {{D.
  Zhu}}, \citenamefont {{S. Song}}, \citenamefont {{J.M. Glownia}},\ and\
  \citenamefont {{S.L. Johnson}}}]{Dornes2019}%
  \BibitemOpen
  \bibfield  {author} {\bibinfo {author} {\bibfnamefont {C.}~\bibnamefont
  {Dornes}}, \bibinfo {author} {\bibfnamefont {Y.}~\bibnamefont {Acremann}},
  \bibinfo {author} {\bibfnamefont {M.}~\bibnamefont {Savoini}}, \bibinfo
  {author} {\bibfnamefont {M.}~\bibnamefont {Kubli}}, \bibinfo {author}
  {\bibnamefont {{M.J. Neugebauer}}}, \bibinfo {author} {\bibfnamefont
  {E.}~\bibnamefont {Abreu}}, \bibinfo {author} {\bibfnamefont
  {L.}~\bibnamefont {Huber}}, \bibinfo {author} {\bibfnamefont
  {G.}~\bibnamefont {Lantz}}, \bibinfo {author} {\bibnamefont {{C.A.F. Vaz}}},
  \bibinfo {author} {\bibfnamefont {H.}~\bibnamefont {Lemke}}, \bibinfo
  {author} {\bibnamefont {{E.M. Bothschafter}}}, \bibinfo {author}
  {\bibnamefont {{M. Porer}}}, \bibinfo {author} {\bibnamefont {{V.
  Esposito}}}, \bibinfo {author} {\bibnamefont {{L. Rettig}}}, \bibinfo
  {author} {\bibnamefont {{M. Buzzi}}}, \bibinfo {author} {\bibnamefont {{A.
  Alberca}}}, \bibinfo {author} {\bibnamefont {{Y.W. Windsor}}}, \bibinfo
  {author} {\bibnamefont {{P. Beaud}}}, \bibinfo {author} {\bibnamefont {{U.
  Staub}}}, \bibinfo {author} {\bibnamefont {{D. Zhu}}}, \bibinfo {author}
  {\bibnamefont {{S. Song}}}, \bibinfo {author} {\bibnamefont {{J.M.
  Glownia}}},\ and\ \bibinfo {author} {\bibnamefont {{S.L. Johnson}}},\ }\href
  {https://doi.org/10.1038/s41586-018-0822-7} {\bibfield  {journal} {\bibinfo
  {journal} {Nature}\ }\textbf {\bibinfo {volume} {565}},\ \bibinfo {pages}
  {209} (\bibinfo {year} {2019})}\BibitemShut {NoStop}%
\bibitem [{\citenamefont {{S.R. Tauchert}}\ \emph {et~al.}()\citenamefont
  {{S.R. Tauchert}}, \citenamefont {Volkov}, \citenamefont {Ehberger},
  \citenamefont {Kazenwadel}, \citenamefont {Evers}, \citenamefont {Lange},
  \citenamefont {Donges}, \citenamefont {Book}, \citenamefont {Kreuzpaintner},
  \citenamefont {Nowak},\ and\ \citenamefont {{P. Baum}}}]{Tauchert2021}%
  \BibitemOpen
  \bibfield  {author} {\bibinfo {author} {\bibnamefont {{S.R. Tauchert}}},
  \bibinfo {author} {\bibfnamefont {M.}~\bibnamefont {Volkov}}, \bibinfo
  {author} {\bibfnamefont {D.}~\bibnamefont {Ehberger}}, \bibinfo {author}
  {\bibfnamefont {D.}~\bibnamefont {Kazenwadel}}, \bibinfo {author}
  {\bibfnamefont {M.}~\bibnamefont {Evers}}, \bibinfo {author} {\bibfnamefont
  {H.}~\bibnamefont {Lange}}, \bibinfo {author} {\bibfnamefont
  {A.}~\bibnamefont {Donges}}, \bibinfo {author} {\bibfnamefont
  {A.}~\bibnamefont {Book}}, \bibinfo {author} {\bibfnamefont {W.}~\bibnamefont
  {Kreuzpaintner}}, \bibinfo {author} {\bibfnamefont {U.}~\bibnamefont
  {Nowak}},\ and\ \bibinfo {author} {\bibnamefont {{P. Baum}}},\ }\Eprint
  {https://arxiv.org/abs/2106.04189} {arXiv:2106.04189} \BibitemShut {NoStop}%
\bibitem [{\citenamefont {Malinowski}\ \emph {et~al.}(2008)\citenamefont
  {Malinowski}, \citenamefont {{F. Dalla Longa}}, \citenamefont {{J.H.H.
  Rietjens}}, \citenamefont {{P.V. Paluskar}}, \citenamefont {Huijink},
  \citenamefont {{H.J.M. Swagten}},\ and\ \citenamefont
  {Koopmans}}]{malinowski2008control}%
  \BibitemOpen
  \bibfield  {author} {\bibinfo {author} {\bibfnamefont {G.}~\bibnamefont
  {Malinowski}}, \bibinfo {author} {\bibnamefont {{F. Dalla Longa}}}, \bibinfo
  {author} {\bibnamefont {{J.H.H. Rietjens}}}, \bibinfo {author} {\bibnamefont
  {{P.V. Paluskar}}}, \bibinfo {author} {\bibfnamefont {R.}~\bibnamefont
  {Huijink}}, \bibinfo {author} {\bibnamefont {{H.J.M. Swagten}}},\ and\
  \bibinfo {author} {\bibfnamefont {B.}~\bibnamefont {Koopmans}},\ }\href
  {https://doi.org/10.1038/nphys1092} {\bibfield  {journal} {\bibinfo
  {journal} {Nat. Phys.}\ }\textbf {\bibinfo {volume} {4}},\ \bibinfo {pages}
  {855} (\bibinfo {year} {2008})}\BibitemShut {NoStop}%
\bibitem [{\citenamefont {Melnikov}\ \emph {et~al.}(2011)\citenamefont
  {Melnikov}, \citenamefont {Razdolski}, \citenamefont {{T.O. Wehling}},
  \citenamefont {{E.Th. Papaioannou}}, \citenamefont {Roddatis}, \citenamefont
  {Fumagalli}, \citenamefont {Aktsipetrov}, \citenamefont {{A.I.
  Lichtenstein}},\ and\ \citenamefont {Bovensiepen}}]{Melnikov2011}%
  \BibitemOpen
  \bibfield  {author} {\bibinfo {author} {\bibfnamefont {A.}~\bibnamefont
  {Melnikov}}, \bibinfo {author} {\bibfnamefont {I.}~\bibnamefont {Razdolski}},
  \bibinfo {author} {\bibnamefont {{T.O. Wehling}}}, \bibinfo {author}
  {\bibnamefont {{E.Th. Papaioannou}}}, \bibinfo {author} {\bibfnamefont
  {V.}~\bibnamefont {Roddatis}}, \bibinfo {author} {\bibfnamefont
  {P.}~\bibnamefont {Fumagalli}}, \bibinfo {author} {\bibfnamefont
  {O.}~\bibnamefont {Aktsipetrov}}, \bibinfo {author} {\bibnamefont {{A.I.
  Lichtenstein}}},\ and\ \bibinfo {author} {\bibfnamefont {U.}~\bibnamefont
  {Bovensiepen}},\ }\href {https://doi.org/10.1103/PhysRevLett.107.076601}
  {\bibfield  {journal} {\bibinfo  {journal} {Phys. Rev. Lett.}\ }\textbf
  {\bibinfo {volume} {107}},\ \bibinfo {pages} {076601} (\bibinfo {year}
  {2011})}\BibitemShut {NoStop}%
\bibitem [{\citenamefont {Battiato}\ \emph {et~al.}(2012)\citenamefont
  {Battiato}, \citenamefont {Carva},\ and\ \citenamefont {{P.M.
  Oppeneer}}}]{Battiato2012}%
  \BibitemOpen
  \bibfield  {author} {\bibinfo {author} {\bibfnamefont {M.}~\bibnamefont
  {Battiato}}, \bibinfo {author} {\bibfnamefont {K.}~\bibnamefont {Carva}},\
  and\ \bibinfo {author} {\bibnamefont {{P.M. Oppeneer}}},\ }\href
  {https://doi.org/10.1103/PhysRevB.86.024404} {\bibfield  {journal} {\bibinfo
  {journal} {Phys. Rev. B}\ }\textbf {\bibinfo {volume} {86}},\ \bibinfo
  {pages} {024404} (\bibinfo {year} {2012})}\BibitemShut {NoStop}%
\bibitem [{\citenamefont {{G.-M. Choi}}\ \emph {et~al.}(2014)\citenamefont
  {{G.-M. Choi}}, \citenamefont {{B.-C. Min}}, \citenamefont {{K.-J. Lee}},\
  and\ \citenamefont {{D.G. Cahill}}}]{Choi2014}%
  \BibitemOpen
  \bibfield  {author} {\bibinfo {author} {\bibnamefont {{G.-M. Choi}}},
  \bibinfo {author} {\bibnamefont {{B.-C. Min}}}, \bibinfo {author}
  {\bibnamefont {{K.-J. Lee}}},\ and\ \bibinfo {author} {\bibnamefont {{D.G.
  Cahill}}},\ }\href {https://doi.org/10.1038/ncomms5334} {\bibfield  {journal}
  {\bibinfo  {journal} {Nat. Commun.}\ }\textbf {\bibinfo {volume} {5}},\
  \bibinfo {pages} {4334} (\bibinfo {year} {2014})}\BibitemShut {NoStop}%
\bibitem [{\citenamefont {Kampfrath}\ \emph {et~al.}(2013)\citenamefont
  {Kampfrath}, \citenamefont {Battiato}, \citenamefont {Maldonado},
  \citenamefont {Eilers}, \citenamefont {N{\"o}tzold}, \citenamefont
  {M{\"a}hrlein}, \citenamefont {Zbarsky}, \citenamefont {Freimuth},
  \citenamefont {Mokrousov}, \citenamefont {Bl{\"u}gel}, \citenamefont {{M.
  Wolf}}, \citenamefont {{I. Radu}}, \citenamefont {{P.M. Oppeneer}},\ and\
  \citenamefont {{M. M{\"u}nzenberg}}}]{Kampfrath2013}%
  \BibitemOpen
  \bibfield  {author} {\bibinfo {author} {\bibfnamefont {T.}~\bibnamefont
  {Kampfrath}}, \bibinfo {author} {\bibfnamefont {M.}~\bibnamefont {Battiato}},
  \bibinfo {author} {\bibfnamefont {P.}~\bibnamefont {Maldonado}}, \bibinfo
  {author} {\bibfnamefont {G.}~\bibnamefont {Eilers}}, \bibinfo {author}
  {\bibfnamefont {J.}~\bibnamefont {N{\"o}tzold}}, \bibinfo {author}
  {\bibfnamefont {S.}~\bibnamefont {M{\"a}hrlein}}, \bibinfo {author}
  {\bibfnamefont {V.}~\bibnamefont {Zbarsky}}, \bibinfo {author} {\bibfnamefont
  {F.}~\bibnamefont {Freimuth}}, \bibinfo {author} {\bibfnamefont
  {Y.}~\bibnamefont {Mokrousov}}, \bibinfo {author} {\bibfnamefont
  {S.}~\bibnamefont {Bl{\"u}gel}}, \bibinfo {author} {\bibnamefont {{M.
  Wolf}}}, \bibinfo {author} {\bibnamefont {{I. Radu}}}, \bibinfo {author}
  {\bibnamefont {{P.M. Oppeneer}}},\ and\ \bibinfo {author} {\bibnamefont {{M.
  M{\"u}nzenberg}}},\ }\href {https://doi.org/10.1038/nnano.2013.43} {\bibfield
   {journal} {\bibinfo  {journal} {Nature Nanotechnology}\ }\textbf {\bibinfo
  {volume} {8}},\ \bibinfo {pages} {256} (\bibinfo {year} {2013})}\BibitemShut
  {NoStop}%
\bibitem [{\citenamefont {Choi}\ \emph {et~al.}(2015)\citenamefont {Choi},
  \citenamefont {Moon}, \citenamefont {Min}, \citenamefont {Lee},\ and\
  \citenamefont {{D.G. Cahill}}}]{Choi2015}%
  \BibitemOpen
  \bibfield  {author} {\bibinfo {author} {\bibfnamefont {G.-M.}\ \bibnamefont
  {Choi}}, \bibinfo {author} {\bibfnamefont {C.-H.}\ \bibnamefont {Moon}},
  \bibinfo {author} {\bibfnamefont {B.-C.}\ \bibnamefont {Min}}, \bibinfo
  {author} {\bibfnamefont {K.-J.}\ \bibnamefont {Lee}},\ and\ \bibinfo {author}
  {\bibnamefont {{D.G. Cahill}}},\ }\href {https://doi.org/10.1038/nphys3355}
  {\bibfield  {journal} {\bibinfo  {journal} {Nat. Phys.}\ }\textbf {\bibinfo
  {volume} {11}},\ \bibinfo {pages} {576} (\bibinfo {year} {2015})}\BibitemShut
  {NoStop}%
\bibitem [{\citenamefont {Alekhin}\ \emph {et~al.}(2017)\citenamefont
  {Alekhin}, \citenamefont {Razdolski}, \citenamefont {Ilin}, \citenamefont
  {{J.P. Meyburg}}, \citenamefont {Diesing}, \citenamefont {Roddatis},
  \citenamefont {Rungger}, \citenamefont {Stamenova}, \citenamefont {Sanvito},
  \citenamefont {Bovensiepen},\ and\ \citenamefont {Melnikov}}]{Alekhin2017}%
  \BibitemOpen
  \bibfield  {author} {\bibinfo {author} {\bibfnamefont {A.}~\bibnamefont
  {Alekhin}}, \bibinfo {author} {\bibfnamefont {I.}~\bibnamefont {Razdolski}},
  \bibinfo {author} {\bibfnamefont {N.}~\bibnamefont {Ilin}}, \bibinfo {author}
  {\bibnamefont {{J.P. Meyburg}}}, \bibinfo {author} {\bibfnamefont
  {D.}~\bibnamefont {Diesing}}, \bibinfo {author} {\bibfnamefont
  {V.}~\bibnamefont {Roddatis}}, \bibinfo {author} {\bibfnamefont
  {I.}~\bibnamefont {Rungger}}, \bibinfo {author} {\bibfnamefont
  {M.}~\bibnamefont {Stamenova}}, \bibinfo {author} {\bibfnamefont
  {S.}~\bibnamefont {Sanvito}}, \bibinfo {author} {\bibfnamefont
  {U.}~\bibnamefont {Bovensiepen}},\ and\ \bibinfo {author} {\bibfnamefont
  {A.}~\bibnamefont {Melnikov}},\ }\href
  {https://doi.org/10.1103/PhysRevLett.119.017202} {\bibfield  {journal}
  {\bibinfo  {journal} {Phys. Rev. Lett.}\ }\textbf {\bibinfo {volume} {119}},\
  \bibinfo {pages} {017202} (\bibinfo {year} {2017})}\BibitemShut {NoStop}%
\bibitem [{\citenamefont {Kimling}\ and\ \citenamefont {{D.G.
  Cahill}}(2017)}]{Kimling2017}%
  \BibitemOpen
  \bibfield  {author} {\bibinfo {author} {\bibfnamefont {J.}~\bibnamefont
  {Kimling}}\ and\ \bibinfo {author} {\bibnamefont {{D.G. Cahill}}},\ }\href
  {https://doi.org/10.1103/PhysRevB.95.014402} {\bibfield  {journal} {\bibinfo
  {journal} {Phys. Rev. B}\ }\textbf {\bibinfo {volume} {95}},\ \bibinfo
  {pages} {014402} (\bibinfo {year} {2017})}\BibitemShut {NoStop}%
\bibitem [{\citenamefont {{T.S. Seifert}}\ \emph {et~al.}(2018)\citenamefont
  {{T.S. Seifert}}, \citenamefont {Jaiswal}, \citenamefont {Barker},
  \citenamefont {{S.T. Weber}}, \citenamefont {Razdolski}, \citenamefont
  {Cramer}, \citenamefont {Gueckstock}, \citenamefont {{S.F. Maehrlein}},
  \citenamefont {Nadvornik}, \citenamefont {Watanabe}, \citenamefont
  {Ciccarelli}, \citenamefont {Melnikov}, \citenamefont {Jakob}, \citenamefont
  {{M. M{\"u}nzenberg}}, \citenamefont {{S.T.B. Goennenwein}}, \citenamefont
  {Woltersdorf}, \citenamefont {Rethfeld}, \citenamefont {{P.W. Brouwer}},
  \citenamefont {Wolf}, \citenamefont {Kl{\"a}ui},\ and\ \citenamefont
  {Kampfrath}}]{Seifert2018}%
  \BibitemOpen
  \bibfield  {author} {\bibinfo {author} {\bibnamefont {{T.S. Seifert}}},
  \bibinfo {author} {\bibfnamefont {S.}~\bibnamefont {Jaiswal}}, \bibinfo
  {author} {\bibfnamefont {J.}~\bibnamefont {Barker}}, \bibinfo {author}
  {\bibnamefont {{S.T. Weber}}}, \bibinfo {author} {\bibfnamefont
  {I.}~\bibnamefont {Razdolski}}, \bibinfo {author} {\bibfnamefont
  {J.}~\bibnamefont {Cramer}}, \bibinfo {author} {\bibfnamefont
  {O.}~\bibnamefont {Gueckstock}}, \bibinfo {author} {\bibnamefont {{S.F.
  Maehrlein}}}, \bibinfo {author} {\bibfnamefont {L.}~\bibnamefont
  {Nadvornik}}, \bibinfo {author} {\bibfnamefont {S.}~\bibnamefont {Watanabe}},
  \bibinfo {author} {\bibfnamefont {C.}~\bibnamefont {Ciccarelli}}, \bibinfo
  {author} {\bibfnamefont {A.}~\bibnamefont {Melnikov}}, \bibinfo {author}
  {\bibfnamefont {G.}~\bibnamefont {Jakob}}, \bibinfo {author} {\bibnamefont
  {{M. M{\"u}nzenberg}}}, \bibinfo {author} {\bibnamefont {{S.T.B.
  Goennenwein}}}, \bibinfo {author} {\bibfnamefont {G.}~\bibnamefont
  {Woltersdorf}}, \bibinfo {author} {\bibfnamefont {B.}~\bibnamefont
  {Rethfeld}}, \bibinfo {author} {\bibnamefont {{P.W. Brouwer}}}, \bibinfo
  {author} {\bibfnamefont {M.}~\bibnamefont {Wolf}}, \bibinfo {author}
  {\bibfnamefont {M.}~\bibnamefont {Kl{\"a}ui}},\ and\ \bibinfo {author}
  {\bibfnamefont {T.}~\bibnamefont {Kampfrath}},\ }\href
  {https://doi.org/10.1038/s41467-018-05135-2} {\bibfield  {journal} {\bibinfo
  {journal} {Nat. Commun.}\ }\textbf {\bibinfo {volume} {9}},\ \bibinfo {pages}
  {2899} (\bibinfo {year} {2018})}\BibitemShut {NoStop}%
\bibitem [{\citenamefont {Iihama}\ \emph {et~al.}(2018)\citenamefont {Iihama},
  \citenamefont {Xu}, \citenamefont {Deb}, \citenamefont {Malinowski},
  \citenamefont {Hehn}, \citenamefont {Gorchon}, \citenamefont {{E.E.
  Fullerton}},\ and\ \citenamefont {Mangin}}]{Iihama2018}%
  \BibitemOpen
  \bibfield  {author} {\bibinfo {author} {\bibfnamefont {S.}~\bibnamefont
  {Iihama}}, \bibinfo {author} {\bibfnamefont {Y.}~\bibnamefont {Xu}}, \bibinfo
  {author} {\bibfnamefont {M.}~\bibnamefont {Deb}}, \bibinfo {author}
  {\bibfnamefont {G.}~\bibnamefont {Malinowski}}, \bibinfo {author}
  {\bibfnamefont {M.}~\bibnamefont {Hehn}}, \bibinfo {author} {\bibfnamefont
  {J.}~\bibnamefont {Gorchon}}, \bibinfo {author} {\bibnamefont {{E.E.
  Fullerton}}},\ and\ \bibinfo {author} {\bibfnamefont {S.}~\bibnamefont
  {Mangin}},\ }\href {https://doi.org/10.1002/adma.201804004} {\bibfield
  {journal} {\bibinfo  {journal} {Advanced Materials}\ }\textbf {\bibinfo
  {volume} {30}},\ \bibinfo {pages} {1804004} (\bibinfo {year}
  {2018})}\BibitemShut {NoStop}%
\bibitem [{\citenamefont {Igarashi}\ \emph {et~al.}(2020)\citenamefont
  {Igarashi}, \citenamefont {Remy}, \citenamefont {Iihama}, \citenamefont
  {Malinowski}, \citenamefont {Hehn}, \citenamefont {Gorchon}, \citenamefont
  {Hohlfeld}, \citenamefont {Fukami}, \citenamefont {Ohno},\ and\ \citenamefont
  {Mangin}}]{Igarashi2020}%
  \BibitemOpen
  \bibfield  {author} {\bibinfo {author} {\bibfnamefont {J.}~\bibnamefont
  {Igarashi}}, \bibinfo {author} {\bibfnamefont {Q.}~\bibnamefont {Remy}},
  \bibinfo {author} {\bibfnamefont {S.}~\bibnamefont {Iihama}}, \bibinfo
  {author} {\bibfnamefont {G.}~\bibnamefont {Malinowski}}, \bibinfo {author}
  {\bibfnamefont {M.}~\bibnamefont {Hehn}}, \bibinfo {author} {\bibfnamefont
  {J.}~\bibnamefont {Gorchon}}, \bibinfo {author} {\bibfnamefont
  {J.}~\bibnamefont {Hohlfeld}}, \bibinfo {author} {\bibfnamefont
  {S.}~\bibnamefont {Fukami}}, \bibinfo {author} {\bibfnamefont
  {H.}~\bibnamefont {Ohno}},\ and\ \bibinfo {author} {\bibfnamefont
  {S.}~\bibnamefont {Mangin}},\ }\href
  {https://doi.org/10.1021/acs.nanolett.0c03373} {\bibfield  {journal}
  {\bibinfo  {journal} {Nano Letters}\ }\textbf {\bibinfo {volume} {20}},\
  \bibinfo {pages} {8654} (\bibinfo {year} {2020})}\BibitemShut {NoStop}%
\bibitem [{\citenamefont {Remy}\ \emph {et~al.}(2020)\citenamefont {Remy},
  \citenamefont {Igarashi}, \citenamefont {Iihama}, \citenamefont {Malinowski},
  \citenamefont {Hehn}, \citenamefont {Gorchon}, \citenamefont {Hohlfeld},
  \citenamefont {Fukami}, \citenamefont {Ohno},\ and\ \citenamefont
  {Mangin}}]{Remy2020}%
  \BibitemOpen
  \bibfield  {author} {\bibinfo {author} {\bibfnamefont {Q.}~\bibnamefont
  {Remy}}, \bibinfo {author} {\bibfnamefont {J.}~\bibnamefont {Igarashi}},
  \bibinfo {author} {\bibfnamefont {S.}~\bibnamefont {Iihama}}, \bibinfo
  {author} {\bibfnamefont {G.}~\bibnamefont {Malinowski}}, \bibinfo {author}
  {\bibfnamefont {M.}~\bibnamefont {Hehn}}, \bibinfo {author} {\bibfnamefont
  {J.}~\bibnamefont {Gorchon}}, \bibinfo {author} {\bibfnamefont
  {J.}~\bibnamefont {Hohlfeld}}, \bibinfo {author} {\bibfnamefont
  {S.}~\bibnamefont {Fukami}}, \bibinfo {author} {\bibfnamefont
  {H.}~\bibnamefont {Ohno}},\ and\ \bibinfo {author} {\bibfnamefont
  {S.}~\bibnamefont {Mangin}},\ }\href {https://doi.org/10.1002/advs.202001996}
  {\bibfield  {journal} {\bibinfo  {journal} {Advanced Science}\ }\textbf
  {\bibinfo {volume} {7}},\ \bibinfo {pages} {2001996} (\bibinfo {year}
  {2020})}\BibitemShut {NoStop}%
\bibitem [{\citenamefont {Beens}\ \emph {et~al.}(2020)\citenamefont {Beens},
  \citenamefont {{R.A. Duine}},\ and\ \citenamefont {Koopmans}}]{Beens2020}%
  \BibitemOpen
  \bibfield  {author} {\bibinfo {author} {\bibfnamefont {M.}~\bibnamefont
  {Beens}}, \bibinfo {author} {\bibnamefont {{R.A. Duine}}},\ and\ \bibinfo
  {author} {\bibfnamefont {B.}~\bibnamefont {Koopmans}},\ }\href
  {https://doi.org/10.1103/PhysRevB.102.054442} {\bibfield  {journal} {\bibinfo
   {journal} {Phys. Rev. B}\ }\textbf {\bibinfo {volume} {102}},\ \bibinfo
  {pages} {054442} (\bibinfo {year} {2020})}\BibitemShut {NoStop}%
\bibitem [{\citenamefont {{Y.L.W. van Hees}}\ \emph {et~al.}(2020)\citenamefont
  {{Y.L.W. van Hees}}, \citenamefont {van~de Meugheuvel}, \citenamefont
  {Koopmans},\ and\ \citenamefont {Lavrijsen}}]{VanHees2020}%
  \BibitemOpen
  \bibfield  {author} {\bibinfo {author} {\bibnamefont {{Y.L.W. van Hees}}},
  \bibinfo {author} {\bibfnamefont {P.}~\bibnamefont {van~de Meugheuvel}},
  \bibinfo {author} {\bibfnamefont {B.}~\bibnamefont {Koopmans}},\ and\
  \bibinfo {author} {\bibfnamefont {R.}~\bibnamefont {Lavrijsen}},\ }\href
  {https://doi.org/10.1038/s41467-020-17676-6} {\bibfield  {journal} {\bibinfo
  {journal} {Nat. Commun.}\ }\textbf {\bibinfo {volume} {11}},\ \bibinfo
  {pages} {3835} (\bibinfo {year} {2020})}\BibitemShut {NoStop}%
\bibitem [{\citenamefont {Iihama}\ \emph {et~al.}(2021)\citenamefont {Iihama},
  \citenamefont {Remy}, \citenamefont {Igarashi}, \citenamefont {Malinowski},
  \citenamefont {Hehn},\ and\ \citenamefont {Mangin}}]{Iihama2021}%
  \BibitemOpen
  \bibfield  {author} {\bibinfo {author} {\bibfnamefont {S.}~\bibnamefont
  {Iihama}}, \bibinfo {author} {\bibfnamefont {Q.}~\bibnamefont {Remy}},
  \bibinfo {author} {\bibfnamefont {J.}~\bibnamefont {Igarashi}}, \bibinfo
  {author} {\bibfnamefont {G.}~\bibnamefont {Malinowski}}, \bibinfo {author}
  {\bibfnamefont {M.}~\bibnamefont {Hehn}},\ and\ \bibinfo {author}
  {\bibfnamefont {S.}~\bibnamefont {Mangin}},\ }\href
  {https://doi.org/10.7566/JPSJ.90.081009} {\bibfield  {journal} {\bibinfo
  {journal} {Journal of the Physical Society of Japan}\ }\textbf {\bibinfo
  {volume} {90}},\ \bibinfo {pages} {081009} (\bibinfo {year}
  {2021})}\BibitemShut {NoStop}%
\bibitem [{\citenamefont {Lichtenberg}\ \emph {et~al.}(2022)\citenamefont
  {Lichtenberg}, \citenamefont {Beens}, \citenamefont {{M.H. Jansen}},
  \citenamefont {{R.A. Duine}},\ and\ \citenamefont
  {Koopmans}}]{Lichtenberg2022}%
  \BibitemOpen
  \bibfield  {author} {\bibinfo {author} {\bibfnamefont {T.}~\bibnamefont
  {Lichtenberg}}, \bibinfo {author} {\bibfnamefont {M.}~\bibnamefont {Beens}},
  \bibinfo {author} {\bibnamefont {{M.H. Jansen}}}, \bibinfo {author}
  {\bibnamefont {{R.A. Duine}}},\ and\ \bibinfo {author} {\bibfnamefont
  {B.}~\bibnamefont {Koopmans}},\ }\href
  {https://doi.org/10.1103/PhysRevB.105.144416} {\bibfield  {journal} {\bibinfo
   {journal} {Phys. Rev. B}\ }\textbf {\bibinfo {volume} {105}},\ \bibinfo
  {pages} {144416} (\bibinfo {year} {2022})}\BibitemShut {NoStop}%
\bibitem [{\citenamefont {{S.M. Rouzegar}}\ \emph {et~al.}()\citenamefont
  {{S.M. Rouzegar}}, \citenamefont {Brandt}, \citenamefont {Nadvornik},
  \citenamefont {{D.A. Reiss}}, \citenamefont {{A.L. Chekhov}}, \citenamefont
  {Gueckstock}, \citenamefont {In}, \citenamefont {Wolf}, \citenamefont {{T.S.
  Seifert}}, \citenamefont {{P.W. Brouwer}}, \citenamefont {{G. Woltersdorf}},\
  and\ \citenamefont {{T. Kampfrath}}}]{Rouzegar2021}%
  \BibitemOpen
  \bibfield  {author} {\bibinfo {author} {\bibnamefont {{S.M. Rouzegar}}},
  \bibinfo {author} {\bibfnamefont {L.}~\bibnamefont {Brandt}}, \bibinfo
  {author} {\bibfnamefont {L.}~\bibnamefont {Nadvornik}}, \bibinfo {author}
  {\bibnamefont {{D.A. Reiss}}}, \bibinfo {author} {\bibnamefont {{A.L.
  Chekhov}}}, \bibinfo {author} {\bibfnamefont {O.}~\bibnamefont {Gueckstock}},
  \bibinfo {author} {\bibfnamefont {C.}~\bibnamefont {In}}, \bibinfo {author}
  {\bibfnamefont {M.}~\bibnamefont {Wolf}}, \bibinfo {author} {\bibnamefont
  {{T.S. Seifert}}}, \bibinfo {author} {\bibnamefont {{P.W. Brouwer}}},
  \bibinfo {author} {\bibnamefont {{G. Woltersdorf}}},\ and\ \bibinfo {author}
  {\bibnamefont {{T. Kampfrath}}},\ }\Eprint {https://arxiv.org/abs/2103.11710}
  {arXiv:2103.11710} \BibitemShut {NoStop}%
\bibitem [{\citenamefont {Shin}\ \emph {et~al.}(2018)\citenamefont {Shin},
  \citenamefont {Min}, \citenamefont {Ju},\ and\ \citenamefont
  {Choi}}]{Shin2018}%
  \BibitemOpen
  \bibfield  {author} {\bibinfo {author} {\bibfnamefont {I.-H.}\ \bibnamefont
  {Shin}}, \bibinfo {author} {\bibfnamefont {B.-C.}\ \bibnamefont {Min}},
  \bibinfo {author} {\bibfnamefont {B.-K.}\ \bibnamefont {Ju}},\ and\ \bibinfo
  {author} {\bibfnamefont {G.-M.}\ \bibnamefont {Choi}},\ }\href
  {https://doi.org/10.7567/JJAP.57.090307} {\bibfield  {journal} {\bibinfo
  {journal} {Japanese Journal of Applied Physics}\ }\textbf {\bibinfo {volume}
  {57}},\ \bibinfo {pages} {090307} (\bibinfo {year} {2018})}\BibitemShut
  {NoStop}%
\bibitem [{\citenamefont {Tserkovnyak}\ \emph {et~al.}(2005)\citenamefont
  {Tserkovnyak}, \citenamefont {Brataas}, \citenamefont {{G.E.W. Bauer}},\ and\
  \citenamefont {{B.I. Halperin}}}]{tserkovnyak2005nonlocal}%
  \BibitemOpen
  \bibfield  {author} {\bibinfo {author} {\bibfnamefont {Y.}~\bibnamefont
  {Tserkovnyak}}, \bibinfo {author} {\bibfnamefont {A.}~\bibnamefont
  {Brataas}}, \bibinfo {author} {\bibnamefont {{G.E.W. Bauer}}},\ and\ \bibinfo
  {author} {\bibnamefont {{B.I. Halperin}}},\ }\href
  {https://doi.org/10.1103/RevModPhys.77.1375} {\bibfield  {journal} {\bibinfo
  {journal} {Reviews of Modern Physics}\ }\textbf {\bibinfo {volume} {77}},\
  \bibinfo {pages} {1375} (\bibinfo {year} {2005})}\BibitemShut {NoStop}%
\bibitem [{Note1()}]{Note1}%
  \BibitemOpen
  \bibinfo {note} {It should be noted that we keep the spin-flip scattering
  rate $\tau _s$ fixed, meaning that the limit $d\ll \lambda _\protect \mathbf
  {sf}$ actually corresponds to assuming a very large conductivity. The
  expressions presented here are equivalent to the similar calculation in Ref.\
  \cite {Beens2022} for $d\ll \lambda _\protect \mathrm {sf}$.}\BibitemShut
  {Stop}%
\bibitem [{\citenamefont {{M. Beens}}\ \emph {et~al.}(2022)\citenamefont {{M.
  Beens}}, \citenamefont {{R.A. Duine}},\ and\ \citenamefont {{B.
  Koopmans}}}]{Beens2022}%
  \BibitemOpen
  \bibfield  {author} {\bibinfo {author} {\bibnamefont {{M. Beens}}}, \bibinfo
  {author} {\bibnamefont {{R.A. Duine}}},\ and\ \bibinfo {author} {\bibnamefont
  {{B. Koopmans}}},\ }\href {https://doi.org/10.1103/PhysRevB.105.144420}
  {\bibfield  {journal} {\bibinfo  {journal} {Phys. Rev. B}\ }\textbf {\bibinfo
  {volume} {105}},\ \bibinfo {pages} {144420} (\bibinfo {year}
  {2022})}\BibitemShut {NoStop}%
\bibitem [{\citenamefont {{S.A. Bender}}\ \emph {et~al.}(2012)\citenamefont
  {{S.A. Bender}}, \citenamefont {{R.A. Duine}},\ and\ \citenamefont
  {Tserkovnyak}}]{Bender2012}%
  \BibitemOpen
  \bibfield  {author} {\bibinfo {author} {\bibnamefont {{S.A. Bender}}},
  \bibinfo {author} {\bibnamefont {{R.A. Duine}}},\ and\ \bibinfo {author}
  {\bibfnamefont {Y.}~\bibnamefont {Tserkovnyak}},\ }\href
  {https://doi.org/10.1103/PhysRevLett.108.246601} {\bibfield  {journal}
  {\bibinfo  {journal} {Phys. Rev. Lett.}\ }\textbf {\bibinfo {volume} {108}},\
  \bibinfo {pages} {246601} (\bibinfo {year} {2012})}\BibitemShut {NoStop}%
\bibitem [{\citenamefont {{N.W. Ashcroft}}\ and\ \citenamefont {{N.D.
  Mermin}}(1976)}]{Ashcroft1976}%
  \BibitemOpen
  \bibfield  {author} {\bibinfo {author} {\bibnamefont {{N.W. Ashcroft}}}\ and\
  \bibinfo {author} {\bibnamefont {{N.D. Mermin}}},\ }\href@noop {} {\emph
  {\bibinfo {title} {Solid State Physics}}}\ (\bibinfo  {publisher} {Holt,
  Rinehart and Winston},\ \bibinfo {year} {1976})\BibitemShut {NoStop}%
\bibitem [{\citenamefont {Kang}\ and\ \citenamefont {Choi}(2020)}]{Kang2020}%
  \BibitemOpen
  \bibfield  {author} {\bibinfo {author} {\bibfnamefont {K.}~\bibnamefont
  {Kang}}\ and\ \bibinfo {author} {\bibfnamefont {G.-M.}\ \bibnamefont
  {Choi}},\ }\href {https://doi.org/10.1016/j.jmmm.2020.167156} {\bibfield
  {journal} {\bibinfo  {journal} {Journal of Magnetism and Magnetic Materials}\
  }\textbf {\bibinfo {volume} {514}},\ \bibinfo {pages} {167156} (\bibinfo
  {year} {2020})}\BibitemShut {NoStop}%
\end{thebibliography}
\providecommand{\noopsort}[1]{}\providecommand{\singleletter}[1]{#1}%

\appendix
\setcounter{table}{0}
\renewcommand{\thetable}{A\arabic{table}}

\bigskip
\section{Notes on the spin current driven by spin pumping}
\label{sec:app5-A}

In this appendix, we present some details regarding the temporal profile of the spin current induced by spin pumping. To calculate $j_{t,s}^\mathrm{sd}$, we have to perform an inverse Fourier transformation of the right-hand side of Eq.\ (\ref{eq:5-15}),  including the function 
\begin{eqnarray}
\label{eq:5-23}
G(\omega) &=& 
\dfrac{i\omega \tau_e}{(1+i\omega\tau_e)(1+i\omega \tau_m) }.
\end{eqnarray}
Disregarding the factors of $2\pi $ (which in the end all vanish), the function in the time domain is given by
\begin{eqnarray}
\label{eq:5-24}
 G(t) &=& 
\dfrac{\tau_e}{\tau_e-\tau_m} 
\bigg(
\dfrac{e^{-t/\tau_m}}{\tau_m}
-\dfrac{e^{-t/\tau_e}}{\tau_e} 
\bigg)\theta(t)  .
\end{eqnarray}

\begin{figure}[b]
\includegraphics[scale=0.95]{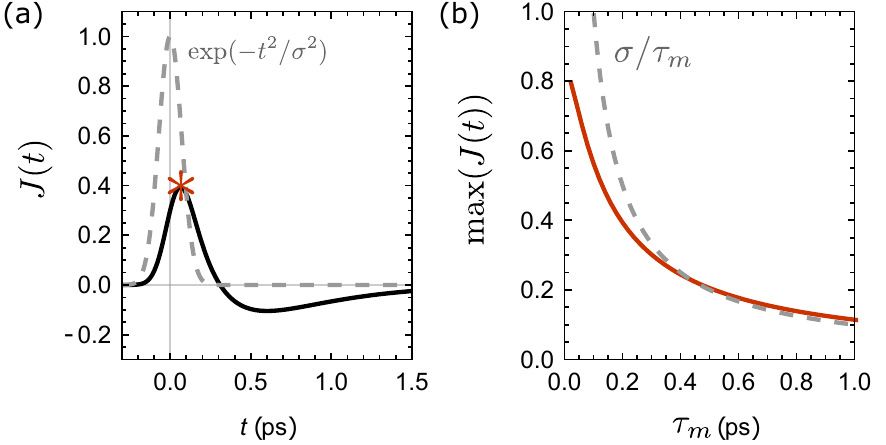}
\caption{\label{fig:5-4} (a) Function $J(t)$ (solid black line) corresponds to the convolution of a Gaussian (dashed gray line) and the function $G(t)$ in the text. (b) The maximum of $J(t)$ plotted as a function of $\tau_m$, compared to the function $\sigma/\tau_m$.    
} 
\end{figure}

The spin current is calculated by performing a convolution between $G(t)$ and the temporal profile of the laser pulse. To determine the scaling factor arising from this convolution we calculate 

\begin{eqnarray}
\label{eq:5-25} 
J(t) &=& 
\int dt' G(t-t')
\exp(-\dfrac{t'^2}{\sigma^2} ).
\end{eqnarray}

The extra scaling factor that should be added to Eq.\ (\ref{eq:5-19})  is given by the maximum of $J(t)$, as it corresponds to how much the (Gaussian) amplitude decreases after the convolution with $G(t)$ is performed. $J(t)$ is plotted in Fig.\ \ref{fig:5-4}(a), together with the temporal profile of the Gaussian pulse. Here we used the values for $\sigma$, $\tau_m$ and $\tau_e$ as given in Table \ref{tab:5-1}. In the range $0.1\mbox{ ps}<\tau_m\leq 1 \mbox{ ps}$, which is the typical order of magnitude for the demagnetization time of a ferromagnetic transition metal, the amplitude scales as $\sigma/\tau_m$, as was mentioned in the main text. The ratio $\sigma/\tau_m$ is indicated by the dashed gray line in Fig.\ \ref{fig:5-4}(b). Finally, we note that Table \ref{tab:5-1} presents the parameters used in the calculations. 

\begin{table}[b]
\centering
\caption{\label{tab:5-1} Parameters used in the calculations presented in the main text.  The chosen values represent a typical magnetic heterostructure consisting of transition metal ferromagnet and a nonmagnetic metal similar to Pt. }
\begin{tabular}[b]{lcc}
\hline \\ [-2.0ex] 
symbol &  value & units  \\
\hline \\ [-1.5ex] 
$\gamma =C_e/T_\mathrm{amb} $ \cite{Kang2020} & $1077$ & $\mbox{Jm}^{-3}\mbox{K}^{-2} $ \\
$T_\mathrm{amb} $  & $300$ & $\mbox{K} $ \\
$\Delta E $  & 1  & $\mbox{eV}$  \\ 
$P_0 $  & $0.2\cdot10^8$  & $\mbox{Jm}^{-3} $ \\ 
$\sigma  $  & $0.1$  & $\mbox{ps} $ \\
$A $ \footnotemark[1]  & $400$ & $\mbox{meV\AA}^{2}$  \\
$d $   & $3$ & $\mbox{nm}$  \\
$\lambda  $ \footnotemark[2]   & $10 $ & $\mbox{nm}$  \\
$P_A $  \footnotemark[3]   & $-0.2$ &   \\
$P_\lambda$ \footnotemark[3]    & $-0.2$ &    \\
$\tau_e $ \cite{Koopmans2005}  & $0.45$ & $\mbox{ps}$  \\
$\tau_m $ \cite{Koopmans2005} & $0.15 $ & $\mbox{ps}$  \\
$\tau_s $ \cite{Shin2018} & $0.1 $ & $\mbox{ps}$  \\
$\tau_g $ \footnotemark[4]    & $0.05  $ & $\mbox{ps}$  \\
\hline
\footnotetext{Used to calculate $C_{n,T}$ as given in Ref. \cite{Beens2022}.}  
\footnotetext{From a decay rate of $~10\mbox{ fs}$ and a Fermi velocity of $~1 \mbox{ nm\, fs}^{-1}$.} 
\footnotetext{A minus sign is present since we defined the spin down electrons as the majority spin population.} 
\footnotetext{Estimated using the values for $g$ (Ni/Pt) and $\tilde{\nu_F}$ (Ni) from Ref.\ \cite{Beens2022}. }
\end{tabular}
\end{table}

\end{document}